\documentclass[my]{myour}
\usepackage{latexsym}
\usepackage{amssymb}
\usepackage{amsmath}
\usepackage{graphics}
\usepackage{epsf}
\usepackage[dvips]{graphicx}
\begin{document}
\title{Event shapes for hadronic final state: experimental review}

\author{V.A. Okorokov\inst{}
\thanks{e-mail: VAOkorokov@mephi.ru;~Okorokov@bnl.gov}%
} 
\institute{National Research Nuclear University MEPhI,
\\ Kashirskoe Shosse 31, 115409 Moscow, Russian Federation}
\date{date May 24, 2011}
%
\abstract{Analysis is presented for first moments of collective
observable distributions in two-jet events for various interaction
types and for wide initial energy range. These observables include
sphericity, thrust, components of transverse particle momentum,
alignment and planarity. Database of experimental results created
in the framework of the paper includes data for all interactions.
Energy dependencies of average values for collective observables
excepting of transverse momentum components show universal
behavior for various interactions. Thus energy dependence for
these parameters supposes the descriptions by functions which are
universal for different interaction types. Particle transverse
momentum as well as it's components increase faster for
$e^{+}e^{-}$ interaction with growth of $\sqrt{s}$, than that for
other interactions. Empirical analytical functions are suggested
for description of energy dependence for all collective
observables under study with exception of infrared-stable thrust
variable. Energy dependence for average thrust is compared with
QCD predictions included perturbative part and analytical
phenomenological corrections which account for non-perturbaive
effects. Dispersive model and single dressed gluon approximation
are considered for description of energy dependence of first
moment of thrust distribution and estimation of strong coupling
constant for various interactions as well as for joint sample. The
dispersive model allow to describe average thrust vs initial
energy in wide range of $\sqrt{s}$ down to strongly
non-perturbative domain $\sqrt{s} \sim 2-3$ GeV at qualitative
level at least. Study of event shape observables allows to obtain
estimations of $\alpha_{S}(M_{Z})$ which are in reasonable
agreement both with world average value and with results extracted
in the framework of other methods. Using suggested analytic
approximation functions estimations of values of collective
parameters under study have been obtained for present and future
facilities. In TeV energy domain average values of collective
observables either depend on $\sqrt{s}$ weakly or do not depend on
initial energy at all within errors. Thus, the TeV scale can be
considered as an estimation of the low bound of asymptotic region
for traditional collective parameters. Usually, multiplicity
dependence of collective observables under considered agree with
power function in energy domain $\sqrt{s} < 12$ GeV at qualitative
level at least. Behavior of sphericity vs multiplicity and
comparison of experimental results with model calculations allow
to suggest that the universal estimation of the low energy
boundary for experimental appearance of event jet structure in
multiparticle production processes is $\sqrt{s_{c}} \sim 3$ GeV.
} 
\maketitle
\section{Introduction}\label{intro}
\hspace*{0.5cm}In various interactions and in wide initial energy
domain final state hadrons appear predominantly in collimated
bunches heading in roughly the same direction within a certain
opening angle. These bunches are generically called \textit{jets}.
Hadronic jets is one of the most spectacular phenomena in particle
physics.

Since we are not able to directly measure colored objects (quarks
or gluons), but only hadrons and their decay products, a central
issue for every experimental test of quantum chromodynamics (QCD)
is establishing a correspondence between observables obtained at
the partonic and the hadronic level. Hadronic jets provide a
window into the fundamental workings of QCD, since they contain
wi\-thin themselves the signatures of QCD at both weak and strong
coupling. Jets therefore test our understanding of QCD over a wide
range of scales [1]. Jets are used also for identifying the hard
partonic structure of decays of massive particles like top quarks.
In order to map observed hadrons onto a set of jets, one uses a
jet definition. Unfortunately formulation of universal and
quantitative definition of hadronic jet for various experiments is
a difficult problem. To a first approximation at enough high
energies, a jet can be thought of as a hard parton that has
undergone soft and collinear showering and then hadronization.
Good jet definitions have to be infrared and collinear safe,
simple to use in theoretical and experimental contexts, applicable
to any type of inputs (parton or hadron momenta, charged particle
tracks, and/or energy deposits in the detectors) and lead to jets
that are not too sensitive to non-perturbative effects [2]. The
review of investigations of jet definitions is given in [3] (for
$e^{+}e^{-}$ annihilation) and in [4, 5] (for hadronic --
$\mbox{pp}$ / $\bar{\mbox{p}}\mbox{p}$ -- collisions).

On the one hand, one can study observables that depend on
specifying the actual number of jets in the final state by
defining a jet algorithm, which sets criteria for what constitutes
a jet. There are two main classes of such algoritms: cone
algorithms, extensively used at hadron colliders, and sequential
recombination algorithms, more widespread at $e^{+}e^{-}$ and
$e\mbox{p}$ colliders. On the other hand, one can extract much
useful information about the structure of the final state from
simpler observables called \textit{event shapes} or collective
variables, which do not depend on a jet algorithm, but are simple
functions of the momenta of all final state particles included in
jet analysis.

Event shapes are used for many purposes. Some collective
characteristics can be calculated perturbatively in terms of QCD
colored objects and compared with hadronic final state
measurements at high energies [6]. The LEP experiments and their
high quality data have prompted much theoretical progress in the
understanding of QCD radiation and in the development of
appropriate tools. This includes Sudakov resummation, renormalon
resummation and parametrization of power corrections [7]. At the
same time, experimental uncertainties and non-per\-tur\-ba\-ti\-ve
effects are small in the regime of high transverse momenta, which
makes possible precision tests of perturbative QCD and the
extraction of the strong coupling constant. On the other hand,
collective observables are directly sensitive to the hadronization
process, providing an opportunity to learn more about the
confinement mechanism from experiment [7]. Understanding the
nature of hadronisation is a challenge, which is crucial important
for all high energy experiments involving hadrons [8]. Thus the
study of collective observables in hadronic collisions and for
wide energy domain seems the important task and promising tool for
investigation of problem of confinement, hadronization mechanism
and corresponding analytical models. At the same time that
hadronic jets provide insights into QCD, they are key elements in
signatures for new physics beyond the Standard Model. Some
traditional event shape observables can be used for search of such
signatures. The detail investigations of the sphericity and thrust
distributions are promising tools for sharp and robust
discrimination between the SUSY and some models with large extra
dimensions which should be utilisable either at ILC or at CLIC
wherever these scenarios become accessible [9]. At a hadronic
colliders such as the LHC, most interesting processes involve
final states with jets of hadrons. The data available to define
and reconstruct the jets are energy depositions in the calorimeter
and charged particle tracks. Ideally one can use this experimental
data to identify an underlying hard scattering event that involved
QCD partons in the weakly coupled high energy regimes or,
alternatively, heavy standard model particles such as top quarks
or newly produced nonstandard model particles. Distinguishing the
former from the latter is of course crucial to our ability to find
and analyze new physics at the LHC [10]. Thus traditional
collective variables can be useful for search of some exotic
events in hadronic collisions at (ultra)high energies. As expected
ordinary QCD events can be distinguished from events involved
decays of new particles because values of some event shapes will
be higher (or closer to the spherical limit) for latter case than
that in the former one. In particular, higher sphericity of an
event in relation to any other event of interactions described by
the Standard Model is of the general regularities for
multidimensional black hole production at accelerators [11]. But
one need to emphasize that the challenge of calculating QCD
background events in the hadron collider environment of the RHIC,
Tevatron or the LHC remains formidable so far.

This paper focuses at experimental study of event sha\-pes namely.
At present the $\mbox{pp}$ colliders RHIC and LHC as well as
Tevatron allows to study the geometry of final state in new energy
ranges at high precisions. Taking into account importance of study
of hadronic interactions but not only collisions with lepton beams
this paper pais attention, especially, to event shapes in
hadron-hadron and hadron-nuclear reactions.

The paper is organized as follows. In Sec.2, definitions of
collective variables under study are described. Database created
in the framework of the paper is shown and discussed in Sec.3. The
Sec.4 devotes to discussion of energy dependence of collective
variables, comparison it with various approximations and extracted
estimations of strong coupling constant. Section 5 demonstrates a
multiplicity dependence of jet characteristics and in Sec.6 some
final remarks are presented.

\section{Observables in analysis of hadronic final state properties}\label{sec:2}
\hspace*{0.5cm}Event shape variables are functions of the four
momenta in the hadronic final state that characterize the topology
of an event's particle / energy flow. They are sensitive to QCD
radiation (and correspondingly to the strong coupling) insofar as
gluon emission changes the shape of the particle / energy flow
[2]. The relationship of pattern of collective particle / energy
flow with measured parameters (momenta) of final state particles
can be described by momentum tensor $\Lambda^{\alpha\beta}$ which
is defined generally in the following way:
\begin{equation}
\Lambda^{\alpha\beta}=\sum\limits_{i=1}^{N}
w_{i}p^{\,i}_{\alpha}p^{\,i}_{\beta},~~~~~\alpha,~\beta=x,~y,~z.\label{eq:1.Tensor-General}
\end{equation}
Here $N$ is the number of particles separated for collective
analysis in event, $w_{i}$ -- some weight of
$i^{\,\mbox{\footnotesize{th}}}$ particle. Both the collective
flow of particles and energy can be investigated at choosing the
suitable $w_{i}$. One need to emphasize that the form of the
momentum tensor varies for different papers (see, for example, [12
-- 16]).

One of the most important and global properties of hadrons
produced in various collisions in wide energy range is their jet
nature which can be characterized by the some collective
observable. The sphericity $(S)$ and thrust $(T)$ are the most
popular and widely used variables for study of jet behavior of the
secondary particles. The sphericity is defined as the following
[12, 13]:
\begin{equation}
S=\min_{\vec{n}} S(\vec{n})=\frac{3}{2} \min_{\vec{n}}
\frac{\textstyle \sum\limits_{i=1}^{N} p^{\,2}_{\perp i}}
{\textstyle \sum\limits_{i=1}^{N}
\vec{p}^{\:2}_{i}}.\label{eq:2.Spherisity}
\end{equation}
Here $p^{\,2}_{\perp i}$ is the transverse momentum of the
$i^{\,\mbox{\footnotesize{th}}}$ particle with respect to the
sphericity axis, which is chosen in such a way that the sum
$\sum\limits_{i=1}^{N} p_{\perp i}^{\,2}$ with respect to it is
minimal; $\vec{p}_{i}$ is the momentum of the particles in the
center-of-mass system, and the sum extends over all particles
separated for jet analysis. Since sphericity is quadratic in the
momenta it cannot be calculated in perturbation part of the QCD,
however it has remained a popular jet measure because it can be
easily determined [17]. A measure of the event structure that uses
the linear momenta is the thrust [18 -- 20]
\begin{equation}
T=\max_{\vec{n}} T(\vec{n})= \max_{\vec{n}} \frac{\textstyle
\sum\limits_{i=1}^{N} |p_{\,\parallel i}|} {\textstyle
\sum\limits_{i=1}^{N} |\vec{p}_{\,i}|}.\label{eq:3.Thrust}
\end{equation}
The thrust axis being chosen to maximize $\sum\limits_{i=1}^{N}
|p_{\,\parallel i}|$, the sum of the components of the momenta
parallel to the thrust axis of all particles in an event selected
for jet analysis.

Sphericity tensor $S^{\alpha\beta}$ can be derived as special case
of general form (\ref{eq:1.Tensor-General}) at $\forall \,i\!:
w_{i}=\mbox{const}$. Usually constant is chosen equal to unit or
as follow $\mbox{const}^{-1}=\sum\limits_{i=1}^{N}
\vec{p}^{\:2}_{i}$. The present work is used the first case of
constant and the matrix of sphericity tensor is the following:
\begin{equation}
S^{\alpha\beta}=\sum\limits_{i=1}^{N}
p_{\,i}^{\,\alpha}p_{\,i}^{\,\beta},~~~~~\alpha,~\beta=x,~y,~z.\label{eq:4.Tensor-Sphericity}
\end{equation}
Because the sphericity tensor uses the momenta of particles
quadratically, the high-momentum particles in an event will
contribute more strongly to observables derived from this tensor
than to those which use the momenta linearly [21]. To eliminate
false effects which may be caused by leading particles, the
reduced-momentum matrix [22] can be also used to study the
configurations of inelastic collisions.

\begin{figure}[h]
\resizebox{0.5\textwidth}{!}{%
  \includegraphics[width=8.1cm,height=7.0cm]{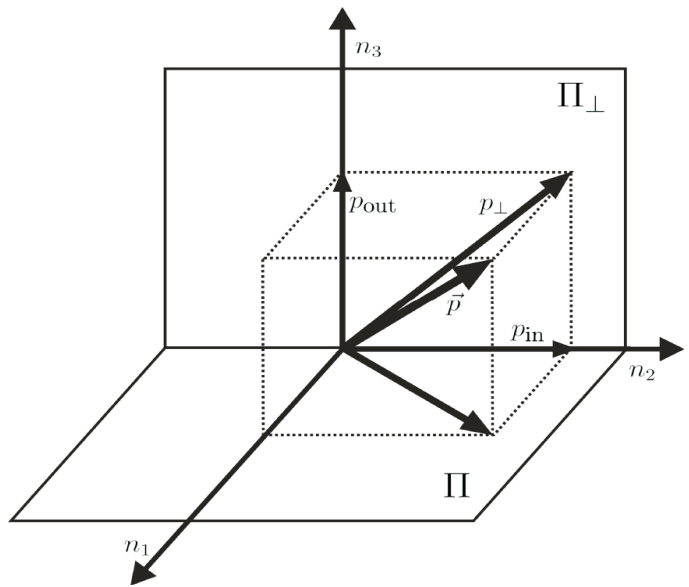}
} \caption{System of coordinates of the "principle axes"
($\vec{n}_{1},~\vec{n}_{2},~\vec{n}_{3}$). Also longitudinal plane
($\Pi$) containing jet axis -- event plane, transverse plane
($\Pi_{\,\perp}$) and components of particle momentum $\vec{p}$ in
transverse plane are shown in the system of coordinate.}
\label{fig:0}
\end{figure}

Study of shape of hadronic final state in various interactions is
carried out in the system of coordinates of the "principle axes"
of the event, which are the symmetry axes for a given
configuration of secondary-particle momentum vectors
(Fig.\ref{fig:0}). This coordinate system has some special
advantages for a more detailed study of the configuration of
events in 3-momentum space. Conversation of this coordinate system
was accomplished by standard procedure of diagonalization of a
matrix (\ref{eq:4.Tensor-Sphericity}). Let
$\vec{n}_{1},~\vec{n}_{2}$, and $\vec{n}_{3}$ be the unit
eigenvectors of the tensor (\ref{eq:4.Tensor-Sphericity})
associated with the eigenvalues $\lambda_{j}$, which are ordered
such that $\lambda_{1} \! \geq \! \lambda_{2} \! \geq \!
\lambda_{3}$ and $\lambda_{\sum} \equiv \sum\limits_{i=1}^{3}
\lambda_{i}$. For this choice the vector $\vec{n}_{1}$ determines
the direction of the greatest extent of the event in momentum
space; $\vec{n}_{2}$ determines the direction of greatest extent
in the plane perpendicular to $\vec{n}_{1}$, and $\vec{n}_{3}$
determines the direction of the greatest compression in this
plane. The axis coincided with direction of eigenvector
$\vec{n}_{1}$ is jet axis in the assumption that the certain event
show two-jet structure. The diagonal elements of the matrix
(\ref{eq:4.Tensor-Sphericity}) represent the sum of the squares of
the components of the secondary-particle momenta in the coordinate
system of the principal axes. If the diagonal elements are related
as $\lambda_{1} \sim \lambda_{2} \sim \lambda_{3}$, then the event
is spherically symmetric; if $\lambda_{1} \gg
\lambda_{2},~\lambda_{1} \gg \lambda_{3}$, and $\lambda_{2} \sim
\lambda_{3}$, the event has cyllindrical symmetry; if $\lambda_{1}
\sim \lambda_{2}$ and $\lambda_{1},~\lambda_{2} \gg \lambda_{3}$,
then the event is planar, recalling a disk in shape [23]. Forms of
events at various values of traditional collective observables are
presented, for example, in [20].

Historically sphericity and thrust are considered as main
traditional variables characterized the geometry of final hadronic
state. There are several additional collective observables derived
from the sphericity tensor. In particular,
$p_{\mbox{\footnotesize{\,in}}}^{\,2}$ and
$p_{\mbox{\footnotesize{\,out}}}^{\,2}$ are, respectively, the
squares of particle transverse momenta {\it in} and {\it out} of
the event plane (Fig.\ref{fig:0}). One of the versions for
definitions of collective variables via eigenvalues of sphericity
tensor $S^{\alpha\beta}$ are shown in the Table \ref{tab:1}.

\begin{table}[!h]
\caption{Event shape observables} \label{tab:1}
\begin{center}
\begin{tabular}{lc} \hline
\multicolumn{1}{c}{Observable} & \multicolumn{1}{c}{Definition} \rule{0pt}{14pt}\\
\hline
sphericity ($S$)  & $3(\lambda_{2}+\lambda_{3}) / 2\lambda_{\footnotesize{\sum}}$ \rule{0pt}{14pt}\\
$p_{\mbox{\footnotesize{\,in}}}^{\,2}$  & $\lambda_{2}
/ N$ \rule{0pt}{14pt}\\
$p_{\mbox{\footnotesize{\,out}}}^{\,2}$  &
$\lambda_{3} / N$ \rule{0pt}{14pt}\\
alignment ($\mbox{A}$) & $\lambda_{2} / \lambda_{1}$ \rule{0pt}{14pt}\\
planarity ($\mbox{P}$)  & $\lambda_{3} / \lambda_{2}$ \rule{0pt}{14pt}\\
\hline
\end{tabular}
\end{center}
\end{table}

According to definitions (\ref{eq:2.Spherisity}) and
(\ref{eq:3.Thrust}) $S \to 1 $ and $T \to 0.5 $ for events with
large number of the secondary particles distributed isotropically
in phase space; $S \to 0 $ and $T \to 1 $ if two separated and
narrow collimated particle groups are produced in interaction
process which are directed to the opposite sides. Magnitudes of
components $p_{\mbox{\footnotesize{\,in}}}^{\,2}$ and
$p_{\mbox{\footnotesize {\,out}}}^{\,2}$ depend on a choice of a
reference plane in a longitudinal direction, i.e. a plane
contained jet axis (plane $\Pi$ at Fig.\ref{fig:0}) while absolute
value of square of particle transverse momentum with respect to
the jet axis
$p_{\footnotesize{\,\perp}}^{\,2}=p_{\mbox{\footnotesize
{\,in}}}^{\,2}+p_{\mbox{\footnotesize {\,out}}}^{\,2}$ does not
depend on a choice of this plane (Fig.\ref{fig:0}). The detailed
review of the experimental results obtained for choice as
(longitudinal) scattering plane the plane formed by the two beams
and the trigger particle is presented, for example, in [24].
Events with greater alignment are characterized by smaller value
of parameter $\mbox{A}$, for strongly planar events is valid the
relationship $\mbox{P} \ll 1$.

Symbol $Y$ denotes below the generic event shape variable from the
set under study. The $n^{\,\mbox{\footnotesize{th}}}$ moment of
the distribution of an event shape parameter $Y$ is defined by
\begin{equation}
\langle Y^{n}\rangle =
\int_{0}^{Y_{\mbox{\scriptsize{max}}}}dYY^{n}\frac{\textstyle
1}{\textstyle \sigma_{\mbox{\footnotesize{tot}}}}\frac{\textstyle
d\sigma}{\textstyle dY},~~n=1, 2, ... \label{eq:GenericCollVar}
\end{equation}
Here $Y_{\mbox{\scriptsize{max}}}$ is the kinematically allowed
upper limit of variable $Y$, $\sigma_{\mbox{\footnotesize{tot}}}$
-- total hadronic cross section. The present paper focuses on the
first moments of distributions of collective variables.

\section{Experimental data sample}\label{sec:3}
\hspace*{0.5cm}Experimental results for event shape variables are
considered for all interaction types and for total available
energy range in the paper. One need to stress that although event
shape of hadronic final state is studied for a long time there is
no standard database for collective characteristics. The
experimental data are presented in the HEPDATA [25] and we made
attempt to collect results for wide set of collective observables.
Table \ref{tab:2} shows corresponding facilities and their
experiments. Results from latest paper of the experiment are
included for given collective parameter at certain energy.

To average data from different experiments at equal or, at least,
(very) close energies weighted procedure is used. We assume that
measurements of given quantity are uncorrelated in different
experiments, and calculate a weighted average as
\begin{equation}
\bar{Y}=\frac{\textstyle \sum_{i}\xi_{i}\langle
Y\rangle_{i}}{\textstyle \sum_{i}\xi_{i}}, \label{eq:4.average}
\end{equation}
where $\langle Y\rangle_{i}$ is the average value of given
observable reported by the $i^{\,\mbox{\footnotesize{th}}}$
experiment, $\xi_{i}$ is the weight of the
$i^{\,\mbox{\footnotesize{th}}}$ experiment which is equal
statistics / luminosity used for study of given observable in the
experiment, and the sums run over the all experiments at certain
energy. Table \ref{tab:3} shows energy values and corresponding
experiments whose results were averaged for some given collective
observable. The notation $\sqrt{s}$ is used for all interaction
types but one need to stress that $Q$ (absolute value of momentum
transfer) and $W$ (invariant mass of final hadronic state) imply
as initial energy parameter for DIS experiments at HERA and for
experiments with (anti)neutrino beams respectively.

It would be stressed that the common basis, in particular,
adequate initial energy scale needs for correct comparison between
various ways of producting multihadronic final states [74].
Identification of the correct initial energy scale, especially,
for hadronic interactions, seems the non-trivial task. There are
different methods for definition of energy scale in hadronic
reactions (see, for example, [60, 63, 66]) but universal scheme is
absent so far. Detail studies in the framework of the paper show
that $\sqrt{s}$ can be used as an energy scale in hadronic
collision together with some specific additional cuts for event /
particle selection (for example, reducing of influence of leading
particle etc.). Comparison of average values of different
collective observables will show reasonable agreement for various
interactions at such choice of energy scale for hadronic
interaction (see below).

Also averaging has been made on Breit variable $x$, beam energy
for some DIS experiments and on different measured kinematic
ranges for some hadronic interactions in order to exclude the
possible influence of additional kinematic cuts on the collective
characteristic values. Errors of experimental points include
available clear indicated systematic errors added in quadrature to
statistical ones.
\begin{table*}
\caption{Facilities and experiments} \label{tab:2}
\begin{center}
\begin{tabular}{lccc} \hline
\multicolumn{1}{c}{Facility} & \multicolumn{1}{c}{Location}
& \multicolumn{1}{c}{Experiment / Detector} & \multicolumn{1}{c}{Refs.} \rule{0pt}{14pt}\\
\hline
\multicolumn{4}{c}{1. $e^{+}e^{-}$ annihilation} \rule{0pt}{14pt}\\
\hline
SPEAR   & SLAC Stanford & SLAC-LBL & ~[13] \rule{0pt}{14pt}\\
PEP     &               & DELCO    & [26] \\
        &               & HRS      & [27] \\
        &               & Mark II  & [28] \\
DORIS   & DESY Hamburg  & PLUTO    & [17, 29]\\
        &               & NaI      & [30]\\
PETRA   &               & CELLO    & [31]\\
        &               & JADE     & [32 -- 34]\\
        &               & MARK-J   & [35, 36]\\
        &               & PLUTO    & [17, 37]\\
        &               & TASSO    & [38, 39]\\
TRISTAN & KEK Tsukuba   & AMY      & [21]\\
        &               & TOPAZ    & [40]\\
SLC     & SLAC Stanford & Mark II  & [41] \\
        &               & SLD      & [42] \\
LEP     & CERN Geneva   & ALEPH    & [43]\\
        &               & DELPHI   & [44]\\
        &               & L3       & [45]\\
        &               & OPAL     & [33, 46]\\
\hline
\multicolumn{4}{c}{2. interactions with lepton beams} \rule{0pt}{14pt}\\
\hline
SPS     & CERN Geneva   & 4m BC (BEBC) & ~[47] \rule{0pt}{14pt}\\
PS      & FNAL Batavia  & 4.6m BC      & [48] \\
HERA    & DESY Hamburg  & H1           & [49 -- 52]\\
        &               & ZEUS         & [53]\\
\hline
\multicolumn{4}{c}{3. hadron-hadron collisions} \rule{0pt}{14pt}\\
\hline
AGS         & BNL Upton     & 2m BC               & ~[54] \rule{0pt}{14pt}\\
PS          & CERN Geneva   & 0.8m and 2m BCs     & [54] \\
            &               & 2m BC               & [22] \\
            &               & 1.5m, 2m, and 4m BCs & [55] \\
            &               & 4m BC               & [56] \\
ISR         &               & CMOR        & [57] \\
            &               & NA5         & [58] \\
            &               & NA23        & [59] \\
            &               & SFM         & [60] \\
S$\bar{\mbox{p}}$pS &       & UA2         & [61] \\
LHC         &               & CMS         & [62] \\
PS          & FNAL Batavia  & IHSC        & [63] \\
            &               & LAFMS       & [64] \\
Tevatron    &               & CDF         & [65] \\
U-70        & IHEP Protvino & 2m BC JINR ("Ludmila")  & [23, 66] \\
            &               & 2m BC       & [67] \\
            &               & 4.5m BC ("Mirabelle") & [68, 69] \\
\hline
\multicolumn{4}{c}{4. hadron-nuclear interactions} \rule{0pt}{14pt}\\
\hline
PS          & ITEP Moscow   & 1m and 2m BCs & ~[70, 71] \rule{0pt}{14pt}\\
U-70        & IHEP Protvino & 2m BC JINR    & [72] \\
SPS         & CERN Geneva   & BEBC          & [47] \\
PS          & FNAL Batavia  & E260          & [73] \\
\hline
\end{tabular}
\end{center}
\end{table*}
One need to emphasize that the most of data for main collective
parameters (sphericity and thrust) are published results. On the
other hand the some values of the above parameters as well as for
other observables are estimations which were obtained on the basis
of mean values of eigenvalues of sphericity tensor or at
experimental distributions of
$p_{\mbox{\footnotesize{\,in}}},~p_{\mbox{\footnotesize{\,out}}}$.
Obviously, such approach allows to get only rough estimations of
true average values. But on the other hand the smooth experimental
dependencies will be presented below and corresponding discussion
show that estimations obtained by this method seem reasonable for
any collective parameters under study and allow to extend energy
domain significantly. Also the values of main collective
observables for special cases as well as, for example, sphericity
and thrust values for quarkonium states [29, 30] or thrust values
in the very narrow energy ranges at searching for top quark [36]
are not shown at the global fits below.

As known definitions of collective observable may vary from one
experiment to other and depend on interaction type and energy
domain under study (for example, there are several definitions for
thrust analogies). But in the framework of this paper we have
tried (if it was possibly) either to estimate the average values
of collective parameter or recalculate they in according with the
definitions shown in Table \ref{tab:1}. Moreover, for thrust we
chose in original papers the definition of the variable which is
most close to the (\ref{eq:3.Thrust}) both in mathematical and
physical meaning for all interactions in high energy domain.
\begin{table*}
\caption{Energy and corresponding experiments for averaging}
\label{tab:3}
\begin{center}
\begin{tabular}{lc} \hline
\multicolumn{1}{c}{$\sqrt{s}$, GeV} &
\multicolumn{1}{c}{Experiments} \rule{0pt}{14pt}\\
\hline
\multicolumn{2}{c}{1. $e^{+}e^{-}$ annihilation} \rule{0pt}{14pt}\\
\hline
9.4            & NaI, PLUTO  \rule{0pt}{14pt}\\
12.0; 13.0; 17.0; 27.3; 30.6 & MARK-J, PLUTO, TASSO \\
14.0           & JADE, TASSO \\
22.0           & JADE, MARK-J, PLUTO, TASSO \\
29.0           & DELCO, HRS, Mark II \\
35.8           & CELLO, JADE, MARK-J, TASSO \\
38.3           & JADE, MARK-J \\
43.6           & CELLO, JADE, TASSO \\
$M_{Z}$        & ALEPH, DELPHI, L3, Mark II, OPAL, SLD \\
133.1 -- 206.1 & ALEPH, DELPHI, L3, OPAL \\
\hline
\multicolumn{2}{c}{2. interactions with lepton beams} \rule{0pt}{14pt}\\
\hline
3.2; 3.9; 5.0; 6.5; 10.6      & BEBC, FNAL BC \rule{0pt}{14pt}\\
9.3; 13.3; 18.3               & H1, ZEUS \\
11.0; 15.2; 17.7; 22.6; 29.7  & \\
36.8; 42.6; 58.9; 82.3; 116.3 & \\
\hline
\end{tabular}
\end{center}
\end{table*}

\section{Energy dependence}\label{sec:4}
\begin{figure*}
\begin{center}
\includegraphics[width=17.5cm]{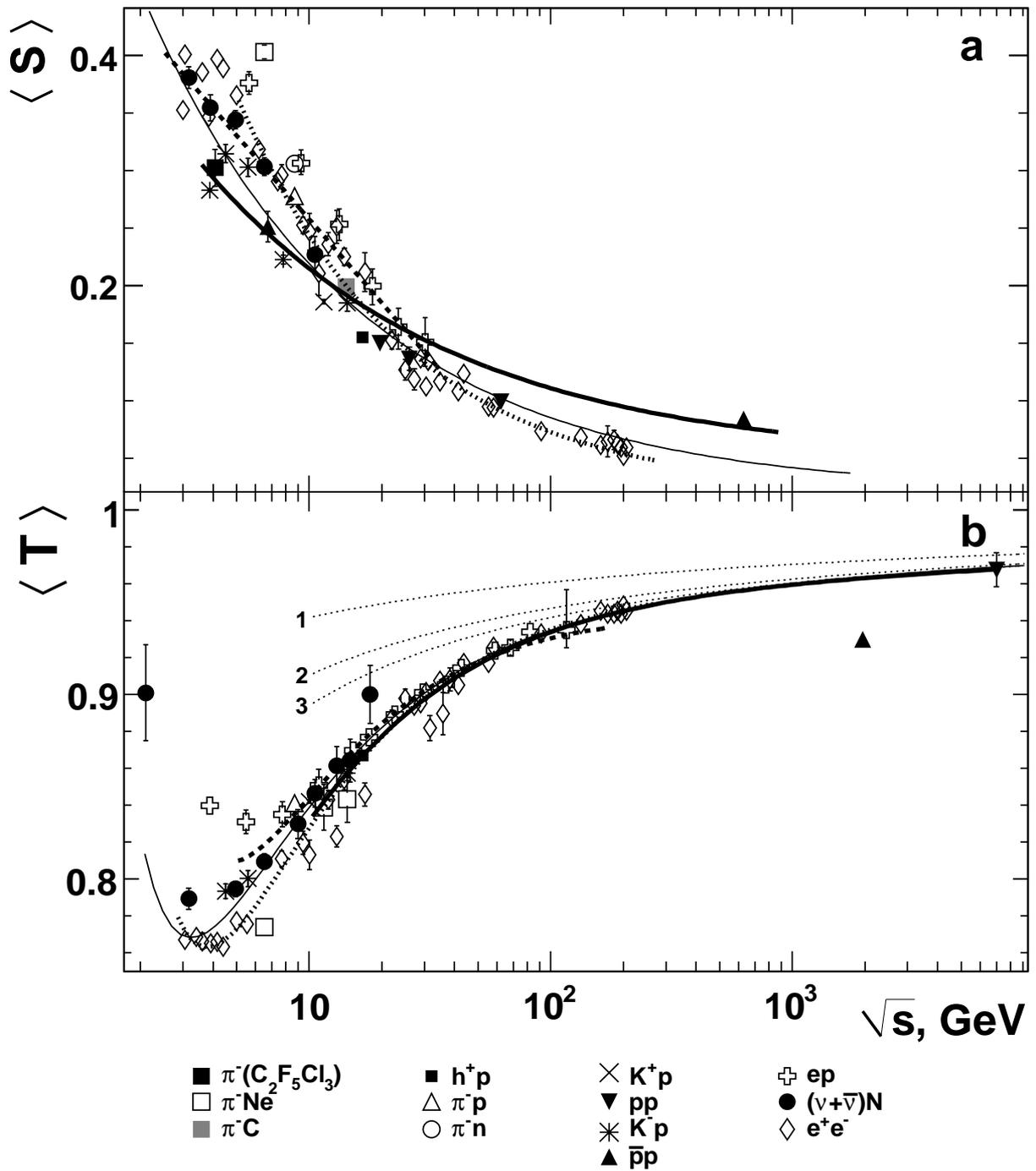}
\end{center}
\caption{Average values of the sphericity (a) and thrust (b)
depend on initial energy. Experimental results and estimations are
from database presented in Table \ref{tab:2}. Total errors are
shown. See text for explanation of the curves.}\label{fig:1}
\end{figure*}

\hspace*{0.5cm}On Fig.\ref{fig:1} dependencies of average values
of collective observable sphericity (a) and thrust (b) on
collision energy are shown. Experimental results and the
estimations for collective characteristics under study are taken
from a database presented in Table \ref{tab:2}. As seen from
Fig.\ref{fig:1} general tendency for interactions of various types
is observed both for sphericity (a) and for thrust (b) at a
qualitative level, moreover the discrepancy of experimental points
reduces at increasing of $\sqrt{s}$.

In the case of hadronic reaction influence of in the leading
particles results in decreasing (increasing) of average value of
$S$ ($T$) in comparison with $e^{+}e^{-}$ and $l\mbox{N}$
collisions. Selection of events with high multiplicity, specific
type of interaction (non-diffractive) or using of modified
sphericity tensor allows to reduce the influence of the leading
particles on jet characteristics in hadronic reactions. Additional
cuts indicated above result in better agreement of values of
$\langle S\rangle$ and $\langle T\rangle$ in various interactions
in the range of intermediate energies. Experimental results for
$\pi^{-}\mbox{Ne}$ collisions at 6.2 GeV/$c$ [70, 71] demonstrate
the influence of nuclear matter on jet observables. Exception of
interactions with single intranuclear nucleon and, probably,
small, in comparison with radius of a target nucleus, length of
forming of final hadronic state [75] leads to essential influence
of multiple rescattering on jet characteristics which results in
wider jets and significantly larger (smaller) average values of
sphericity (thrust) in comparison with results of other
experiments at close energies. The account of interactions with a
single intranuclear nucleon in case of discussed
$\pi^{-}\mbox{Ne}$ interactions leads to $\langle S\rangle=0.269
\pm 0.005$ and $\langle T\rangle=0.814 \pm 0.002$, that would be
agreed with the general tendency much more reasonably. Reducing of
influence of rescattering with increasing of beam momentum leads
to improvement of agreement of $\langle S\rangle$ even for
multinucleon hadron-nuclear interactions with the general tendency
(Fig.\ref{fig:1}a), that is visible on an example of
$\pi^{-}\mbox{C}$ collisions [72]. Thus, integral (on number of
participating nucleons) results confirm a conclusion [71, 72],
that the mechanism of production of hadron jets at nuclei is more
complex, than that in processes with hadrons and / or leptons, and
growth of average values of sphericity for $\mbox{hA}$
interactions at intermediate energies can be explained by multiple
rescattering of secondary particles at nucleons of a target
nucleus. Point for $\pi^{-}\mbox{Ne}$ at initial momentum 6.2
GeV/$c$ discussed above is excluded below from the fitted samples
for sphericity and thrust. At present there are no experimental
results in TeV energy domain for traditional collective
observables which were described above. The experimental results
for transverse thrust ($T_{\perp}$) from [62, 65] are used as
estimations of thrust defined by (\ref{eq:3.Thrust}) at TeV scale
energies. This assumption is based on the relation $T \sim
T_{\perp}$ which valids for events with two back-to-back jets and
for near-to-planar 3-jet events [76, 77, 78]. As seen from
Fig.\ref{fig:1} the CDF experimental result [65] differs
noticeably from general tendency for thrust
observable\footnote{One need to emphasize that the result based on
parton-level NLO+NLL distribution agree much better with general
trend.}, i.e. the assumption is less correct for CDF than that for
CMS. Possibly there are several reasons for such behavior of
experimental point at $\sqrt{s}=1.96$ TeV. As known contribution
of soft particles at large angles with respect to the event plane
can leads to $T \gtrsim T_{\perp}$ [77]. Indeed, CDF data show
slightly intensive energy flow out of the hard-scattering plane
[65]. Thus discrepancy of CDF point from general trend can be
caused, in particular, by more significant contribution of soft
particles at large angles. Additional reason may be kinematical
cuts used in CDF analysis. In any case for verification of this
hypothesis additional detail study is essential for event geometry
with taking into account soft particles, especially, examination
of relation $p_{\mbox{\footnotesize {\,in}}} \sim
p_{\mbox{\footnotesize {\,out}}}$ for such particles. Therefore
the CDF experimental point is excluded below from fitted samples.
\begin{table*}
\caption{Results of fit of dependence $\langle S\rangle (s)$ by
function (\ref{eq:Fit-vs-s-2})} \label{tab:4-sf}
\begin{center}
\begin{tabular}{cccccp{2.0cm}} \hline
\multicolumn{1}{c}{Data sample} & \multicolumn{4}{c}{Fit parameters}\\
\cline{2-5}
 & $a_{1}$ & $a_{2}$ & $a_{3}$ &$\chi^{2}/\mbox{ndf}$ \rule{0pt}{14pt}\\
\hline
$\mbox{hh}$ and $\mbox{hA}$       & $0.047 \pm 0.004$  & $0.442 \pm 0.005$ & $-0.210 \pm 0.006$ & 97.5 \rule{0pt}{14pt}\\
$l\mbox{N}$                       & $-357 \pm 303$     & $358 \pm 303$     & $(-1.5 \pm 1.2) \times 10^{-4}$ & 5.22 \\
$e^{+}e^{-}$ ($\sqrt{s} > 5$ GeV) & $0.019 \pm 0.004$ & $0.94 \pm 0.02$ & $-0.311 \pm 0.009$ & 2.95 \\
all data                          & $0.021 \pm 0.002$  & $0.610 \pm 0.004$ & $-0.244 \pm 0.004$ & 70.0 \\
\hline
\end{tabular}
\end{center}
\end{table*}

There are smooth dependencies of experimental $\langle S\rangle$
on $\sqrt{s}$ for various interactions without qualitative changes
at transition from intermediate energy domain to high energies
(Fig.\ref{fig:1}a). Taking into account general view of sphericity
dependence on initial energy and results for $e^{+}e^{-}$ [27,
39], in the present paper the $\langle S\rangle (s)$ is
approximated by power function
\begin{equation}
\langle Y\rangle=a_{1}+a_{2}X^{a_{3}},\label{eq:Fit-vs-s-2}
\end{equation}
where $Y$ is certain collective variable and $X$ denotes the
appropriate argument of a function, $Y \equiv S$ and $X \equiv s$
in the case under considered. Fit results for various samples of
experimental data for sphericity observable by function
(\ref{eq:Fit-vs-s-2}) are presented in Table \ref{tab:4-sf} and
shown on Fig.\ref{fig:1}a for $\mbox{hh}+\mbox{hA}$ interactions
by a solid line, for $l\mbox{N}$ -- by dashed one, for
$e^{+}e^{-}$ annihilation -- by dotted line and for the joint
sample by thin solid one. The disorder of experimental points,
especially in the range of $\sqrt{s} \leq 5-10$ GeV, does not
allow to obtain statistically acceptable fit quality. Influence of
discrepancy of experimental data from indicated above energy
domain on the approximation quality is confirmed, for example, by
results for $e^{+}e^{-}$ annihilation: for $\sqrt{s} > 5$ GeV
reasonable fit quality is observed (Table \ref{tab:4-sf}),
moreover value of $a_{3}$ parameter agrees within errors with
result from [27], on the other hand, fitting of all available
energy range results in $\chi^{2} / \mbox{ndf} = 19.1$. However
one can see from Fig.\ref{fig:1}, that the behavior of $\langle
S\rangle$ depending on $\sqrt{s}$ is described by function
(\ref{eq:Fit-vs-s-2}) at a qualitative level both for separate
interactions, and for the joint sample for all available range of
initial energies. Additional research for the joint sample with
using of more complex fit function, in particular, the sum of
power functions, has shown, that experimental data are better
described by function (\ref{eq:Fit-vs-s-2}).

Unlike sphericity, collective observable the thrust possesses
infrared stability that allows to use QCD for study of behavior
various dependencies and distributions on this collective
parameter. In the framework of pQCD the energy dependence of the
first moment of thrust distribution can be described by the
following function
\begin{equation}
\langle T\rangle^{\mbox{\scriptsize{pQCD}}}
(s)=1-\sum\limits_{i=1}\bar{\mathcal{A}}_{i}\alpha_{S}^{i}(s),
\quad \alpha_{S}(s)=(b_{0}t)^{-1}. \label{eq:4.4.Fit-vs-s-TpQCD}
\end{equation}
Here $t \equiv \ln s/\Lambda^{2}$, $\Lambda$ is the QCD parameter,
$b_{0}$ is referred to as one-loop $\beta$-function coefficient
[2]. Parameters $\bar{\mathcal{A}}_{i}$ are calculated up to NNLO
in [79] and for (\ref{eq:4.4.Fit-vs-s-TpQCD}) they are
$\bar{\mathcal{A}}_{1}=0.335,~\bar{\mathcal{A}}_{2}=1.033,~\bar{\mathcal{A}}_{3}=2.76
\pm 0.09$ for number of (light, i.e. active) quark flavors
$N_{f}=5$. Fig.\ref{fig:1}b shows the theoretical predictions for
LO ("1"), NLO ("2") and NNLO ("3") pQCD based on the
(\ref{eq:4.4.Fit-vs-s-TpQCD}) by thin dotted curves. As seen
behavior of a theoretical curve $\langle
T\rangle^{\mbox{\scriptsize{pQCD}}} (s)$ depends on the order of
pQCD expansion, especially, in the intermediate energy domain. The
agreement of theoretical prediction (\ref{eq:4.4.Fit-vs-s-TpQCD})
with experimental data improves for higher order of pQCD
expansion. The LO pQCD overpredicts the thrust value even at LHC
energy. It is important to note very good agreement of
experimental results for $\mbox{pp}$ collisions at $\sqrt{s}=7$
TeV [62] and a theoretical curves (\ref{eq:4.4.Fit-vs-s-TpQCD})
for NLO and NNLO. Such quantitative agreement is observed for the
first time, especially, for hadronic reaction and allows to
assume, that at $\sqrt{s} \sim 5-10$ TeV non-perturbative
corrections become negligible. As known the contribution of these
corrections increases at decreasing of initial energy and for
approximation of $\langle T\rangle (s)$ in range of $\sqrt{s}$ as
wide as possible it is necessary to use the phenomenological
models which take into account the non-perturbative corrections.
Thus, in the present work for fit of dependencies $\langle
T\rangle (s)$ in various interactions the following general
function is used:
\begin{equation}
\langle T\rangle (s)=1-[\langle
F\rangle^{\mbox{\scriptsize{pQCD}}}+\langle
F\rangle^{\mbox{\scriptsize{n-p}}}],\label{eq:4.4.Fit-vs-s-Tgeneral}
\end{equation}
where components $\langle F\rangle^{\mbox{\scriptsize{pQCD}}}$ and
$\langle F\rangle^{\mbox{\scriptsize{n-p}}}$ are model dependent.
In the framework of present paper dispersive model (DM) [80, 81]
and single dressed gluon model (SDGM) [82 -- 84] have been used
for approximation of dependence $\langle T\rangle (s)$. It is
necessary to note, that now the most part of calculations are
executed for electron-positron annihilation and comparison of
models with experimental values of collective characteristics is
carried out, as usual, for the this type of interactions because
precision of measurements and calculations for shape variables of
final hadronic state is the best for $e^{+}e^{-}$ namely, in
additional to significant sample of experimental results is
available in the high energy domain where applicability of QCD and
corresponding phenomenological models has more rigorous
substantiation. Some predictions on the basis of DM have been
obtained for $e\mbox{p}$ interactions also [49, 51, 52]. For
hadron-hadron and hadron-nuclear reactions the number of
phenomenological predictions is much less, that is connected both
with the raised difficulty of calculations and with absence of
experimental data at high energies until recently. Therefore the
functional forms obtained for $e^{+}e^{-}$ initially are used
below for fitting of various data samples.

In DM perturbative part is defined in the framework of NLO pQCD
and terms for (\ref{eq:4.4.Fit-vs-s-Tgeneral}) are the following:
\begin{equation}
\begin{split}
&\left\langle
F\right\rangle_{\mbox{\scriptsize{DM}}}^{\mbox{\scriptsize{pQCD}}}=
\sum\limits_{i=1}^{2}
\mathcal{A}_{i}\bar{\alpha}_{S}^{i}(s),~~~~~\bar{\alpha}_{S}(s)
\equiv \alpha_{S}(s) / 2\pi,
\\
&\left\langle F
\right\rangle_{\mbox{\scriptsize{DM}}}^{\mbox{\scriptsize{n-p}}}=\mathcal{M}
\frac{\textstyle 4C_{F}}{\textstyle \pi}\left(\frac{\textstyle
\mu_{I}}{\textstyle \sqrt{s}}\right)\times \\
&\left[\bar{\alpha}_{0}(\mu_{I})-\alpha_{S}(s)-
b_{0}\left(\ln\frac{\textstyle s}{\textstyle
\mu_{I}^{2}}+\frac{\textstyle K}{\textstyle 2\pi
b_{0}}+2\right)\alpha_{S}^{2}(s)\right],\\
&~K=\left(\frac{\textstyle 67}{\textstyle 18}-\frac{\textstyle
\pi^{2}}{\textstyle 6}\right)C_{A}- \frac{\textstyle 5}{\textstyle
9}N_{f}, C_{F}=\frac{\textstyle 4}{\textstyle 3}, C_{A}=3.
\end{split}
\label{eq:4.4.Fit-vs-s-TDM}
\end{equation}
Here $\mathcal{M}$ is correcting so called Milan factor [85, 86].
In this work $\mathcal{M}$ and $\mathcal{A}_{i},~(i=1,2)$ are free
fit parameters. Non-per\-tur\-ba\-ti\-ve parameter
$$
\bar{\alpha}_{p}(\mu_{I})=\frac{\textstyle p+1}{\textstyle
\mu_{I}^{p+1}}\int^{\mu_{I}}_{0} k^{p}\alpha_{S}(k)dk
$$
is introduced for the account of a part of expression for
$\alpha_{S} (s)$ which diverges below $\mu_{I}$ -- infrared
matching scale [80]. There is no unambiguous scheme for definition
of $\mu_{I}$ value. The empirical choice of $\mu_{I}$ and $N_{f}$
values is restricted by the requirement of validity of following
conditions: $\Lambda \ll \mu_{I} < \sqrt{s_{\mbox{\footnotesize
{min}}}}$ [80] and $\forall N_{f}: m_{q} <
\sqrt{s_{\mbox{\footnotesize {min}}}}$ [2], where
$\sqrt{s_{\mbox{\footnotesize{min}}}}$ is the low boundary of
fitted energy range, $m_{q}$ denotes masses of the quarks, which
can be considered as lights for a certain
$\sqrt{s_{\mbox{\footnotesize{min}}}}$. One need to emphasize that
in the framework of present paper, in distinction from, for
example [51], for $l\mbox{N}$ collisions dependencies of
parameters $\mathcal{A}_{i},~(i=1,2)$ on some kinematic quantities
$(x, Q)$ have not been taken into account because, as indicated
above, the averaged on these parameters and various experiments
values of thrust are used. Fit results for various samples of
experimental data for thrust observable by function
(\ref{eq:4.4.Fit-vs-s-Tgeneral}) in the framework of DM are
presented in Table \ref{tab:5-DM} and shown on Fig.\ref{fig:1}b
for $e^{+}e^{-}$ annihilation by dotted curve, for $l\mbox{N}$ --
by dashed one, for $\mbox{hh}+\mbox{hA}$ interactions -- by solid
line, and for the joint sample -- by thin solid one. As usual the
better fit quality is obtained at the fixed value of $\mathcal{M}$
which is defined by limits of the allowable interval of Milan
factor values [85]. As seen from Table \ref{tab:5-DM} and
Fig.\ref{fig:1}b DM allows to describe all available results for
$e^{+}e^{-}$ at reasonable fit quality. The disorder of
experimental points leads to that statistically acceptable values
of $\chi^{2} / \mbox{ndf}$ are observed at
$\sqrt{s_{\mbox{\footnotesize{min}}}}=5$ and 10 GeV only for
$l\mbox{N}$ and hadronic reactions respectively. For these
interactions the fit functions are shown Fig.\ref{fig:1}b at
$\sqrt{s_{\mbox{\footnotesize{min}}}}$ indicated above namely.
However function (\ref{eq:4.4.Fit-vs-s-Tgeneral}) in the framework
of DM allows to describe dependence $\langle T\rangle (s)$ at a
qualitative level both for some interactions and for the joint
data sample in all range of available initial energies
(Fig.\ref{fig:1}b). Behavior of DM-curves agrees very well with
NLO and NNLO pQCD predictions in the LHC energy domain. Thus, for
the first time the reasonable phenomenological description is
obtained for experimental results for thrust variable in all
available energy range both for $e^{+}e^{-}$ and for the joint
data sample for various interactions. Non-perturbative parameter
$\bar{\alpha}_{p}(\mu_{I})$ values obtained for $e^{+}e^{-}$ and
$l\mbox{N}$ at $\sqrt{s_{\mbox{\footnotesize{min}}}}=10$ GeV agree
with previous results [52, 87] within errors. Clear dependence of
$\bar{\alpha}_{p}(\mu_{I})$ on
$\sqrt{s_{\mbox{\footnotesize{min}}}}$ is absent within errors for
these interactions. The similar situation is observed for hadronic
reactions. But one can see growth of $\bar{\alpha}_{p}(\mu_{I})$
value at decreasing of $\sqrt{s_{\mbox{\footnotesize{min}}}}$ for
joint data sample, especially visible at transition in strong
non-perturbative region from
$\sqrt{s_{\mbox{\footnotesize{min}}}}=5$ GeV down to
$\sqrt{s_{\mbox{\footnotesize{min}}}}=2$ GeV (Table
\ref{tab:5-DM}). There is no clear dependence of
$\bar{\alpha}_{p}(\mu_{I})$ value on type of interaction at fixed
low boundary for fitted energy domain.
\begin{table*}
\caption{Results for fit of dependence $\langle T\rangle (s)$ by
function (\ref{eq:4.4.Fit-vs-s-Tgeneral}) in the framework of DM}
\label{tab:5-DM}
\begin{center}
\begin{tabular}{cccccp{2.0cm}} \hline
\multicolumn{1}{c}{Fit} & \multicolumn{4}{c}{Data sample}\\
\cline{2-5}
parameter & $e^{+}e^{-}$ & $l\mbox{N}$ & $\mbox{hh}$ and $\mbox{hA}$ & all data \rule{0pt}{14pt}\\
\hline \multicolumn{5}{c}{$\sqrt{s_{\mbox{\footnotesize{min}}}}=2$
GeV ($\mu_{I}=1$ GeV, $N_{f}=3$)} \rule{0pt}{14pt}\\
\hline
$\mathcal{A}_{1}$            & $3.8 \pm 0.4$       & $2.1 \pm 0.3$       & $4.08 \pm 0.15$     & $2.6 \pm 0.3$ \rule{0pt}{14pt}\\
$\mathcal{A}_{2}$            & $-63 \pm 26$        & $48 \pm 17$         & $-11.1 \pm 0.8$     & $16 \pm 6$ \\
$\mathcal{M}$ (fixed)        & 1.795               & 1.448               & 1.100               & $1.100$ \\
$\bar{\alpha}_{0} (\mu_{I})$ & $0.77 \pm 0.04$     & $0.57 \pm 0.03$     & $0.628 \pm 0.012$   & $0.75 \pm 0.05$\\
$\alpha_{S} (M_{Z})$         & $0.1192 \pm 0.0007$ & $0.1209 \pm 0.0012$ & $0.1180 \pm 0.0011$ & $0.1218 \pm 0.0007$\\
$\chi^{2}/\mbox{ndf}$        & 3.04                & 8.16                & 20.1                & 15.6 \\
\hline \multicolumn{5}{c}{$\sqrt{s_{\mbox{\footnotesize{min}}}}=5$
GeV ($\mu_{I}=2$ GeV, $N_{f}=4$)} \rule{0pt}{14pt}\\
\hline
$\mathcal{A}_{1}$            & $3.17 \pm 0.13$     & $8.02 \pm 0.06$     & $4.2 \pm 1.2$      & $2.76 \pm 0.15$ \rule{0pt}{14pt}\\
$\mathcal{A}_{2}$            & $-32 \pm 4$         & $-300 \pm 7$        & $6 \pm 3$          & $2.7 \pm 1.3$ \\
$\mathcal{M}$ (fixed)        & 1.520               & 1.852               & 1.188              & 1.188 \\
$\bar{\alpha}_{0} (\mu_{I})$ & $0.532 \pm 0.018$   & $0.638 \pm 0.008$   & $0.39 \pm 0.02$    & $0.517 \pm 0.008$\\
$\alpha_{S} (M_{Z})$         & $0.1203 \pm 0.0018$ & $0.1222 \pm 0.0003$ & $0.116 \pm 0.004$  & $0.1206 \pm 0.0009$\\
$\chi^{2}/\mbox{ndf}$        & 3.63                & 1.74                & 20.3               & 7.38 \\
\hline
\multicolumn{5}{c}{$\sqrt{s_{\mbox{\footnotesize{min}}}}=10$ GeV
($\mu_{I}=2$ GeV, $N_{f}=5$)} \rule{0pt}{14pt}\\
\hline
$\mathcal{A}_{1}$            & $4.4 \pm 1.3$       & $-0.6 \pm 0.3$     & $2.7 \pm 1.1$       & $3.7 \pm 1.2$ \rule{0pt}{14pt}\\
$\mathcal{A}_{2}$            & $-97 \pm 66$        & $212 \pm 21$       & $8 \pm 2$           & $-58 \pm 34$ \\
$\mathcal{M}$                & 1.291 (fixed)       & 1.291 (fixed)      & 1.920 (fixed)       & $1.8 \pm 0.5$ \\
$\bar{\alpha}_{0} (\mu_{I})$ & $0.59 \pm 0.18$     & $0.32 \pm 0.17$    & $0.42 \pm 0.10$     & $0.49 \pm 0.04$\\
$\alpha_{S} (M_{Z})$         & $0.115 \pm 0.011$   & $0.115 \pm 0.017$  & $0.117 \pm 0.016$   & $0.121 \pm 0.011$\\
$\chi^{2}/\mbox{ndf}$        & 3.28                & 0.43               & 0.69                & 3.19 \\
\hline
\end{tabular}
\end{center}
\end{table*}

\begin{table*}
\caption{Results for fit of dependence $\langle T\rangle (s)$ by
function (\ref{eq:4.4.Fit-vs-s-Tgeneral}) in the framework of
SDGM} \label{tab:5-SDGM}
\begin{center}
\begin{tabular}{ccccp{2.0cm}} \hline
\multicolumn{1}{c}{Data} & \multicolumn{3}{c}{Fit parameter}\\
\cline{2-4}
sample & $\alpha_{S}(M_{Z})$ & $\nu$ & $\chi^{2}/\mbox{ndf}$ \rule{0pt}{14pt}\\
\hline
$e^{+}e^{-}$ & $0.1235 \pm 0.0008$ & $-0.28 \pm 0.09$   & 6.98 \\
all data     & $0.1197 \pm 0.0005$ & $-0.142 \pm 0.049$ & 14.0 \\
\hline
\end{tabular}
\end{center}
\end{table*}

For $e^{+}e^{-}$ results and for joint data sample approximations
in the framework of SDG-model have been investigated too. In the
case of the first moment of thrust distribution, $\langle
T\rangle$, perturbative and non-perturbative components in SDGM
for (\ref{eq:4.4.Fit-vs-s-Tgeneral}) are the following [87]:
\begin{equation}
\left\langle
F\right\rangle_{\mbox{\scriptsize{SDGM}}}^{\mbox{\scriptsize{pQCD}}}=
\sum\limits_{i=1}^{6} \mathcal{B}_{i}\bar{a}^{i}(s) / \pi,~
\left\langle F
\right\rangle_{\mbox{\scriptsize{SDGM}}}^{\mbox{\scriptsize{n-p}}}=\nu
/ \sqrt{s}. \label{eq:4.4.Fit-vs-s-TSDGM}
\end{equation}
Here
$$
\bar{a}(s) \equiv
\cfrac{\alpha_{S}(s)}{1-\frac{5}{3}b_{0}\alpha_{S}(s)}
$$
and $\nu$ is a free fit parameter. Coefficients $\mathcal{B}_{i}$
are defined via so called log moments $d_{i}$ of the
characteristic thrust function, calculated in [83], by following
way
\begin{equation}
\begin{split}
&\mathcal{B}_{1}=d_{0},~\mathcal{B}_{2}=\beta_{0}d_{1},~
\mathcal{B}_{3}=\left(-\frac{\textstyle d_{0}\pi^{2}}{\textstyle
3}+d_{2}\right)\beta_{0}^{2}+\beta_{1}d_{1}, \\
&\mathcal{B}_{4}=(-\pi^{2}d_{1}+d_{3})\beta_{0}^{3}+
\frac{\textstyle 5}{\textstyle 2}\beta_{1}\beta_{0}d_{2}+\beta_{2}
d_{1}, \\
&\mathcal{B}_{5}=\left(d_{4}+\frac{\textstyle
d_{0}\pi^{4}}{\textstyle 5}-2\pi^{2}d_{2}\right)\beta_{0}^{4}+
\left(\frac{\textstyle 13}{\textstyle
3}\beta_{1}d_{3}-\beta_{1}\pi^{2}d_{1}\right)\beta_{0}^{2}\\
&~~~~~~~+3\beta_{2} d_{2}\beta_{0} +\frac{\textstyle 3}{\textstyle
2}\beta_{1}^{2}d_{2}+\beta_{3}d_{1}, \\
&\mathcal{B}_{6}= \left(-\frac{\textstyle 10}{\textstyle 3}\pi^{2}
d_{3}+\pi^{4}d_{1}+d_{5}\right)\beta_{0}^{5}+
\left(\frac{\textstyle 77}{\textstyle 12}d_{4}-\frac{\textstyle
9}{\textstyle
2}\pi^{2}d_{2}\right) \times\\
&~~~~~~~\beta_{1}\beta_{0}^{3}+(6d_{3}-\pi^{2}d_{1})\beta_{2}
\beta_{0}^{2}+\left(\frac{\textstyle 35}{\textstyle
6}\beta_{1}^{2}d_{3}+\frac{\textstyle 7}{\textstyle
2}\beta_{3}d_{2}\right)\beta_{0}+\\
&~~~~~~~\beta_{4}d_{1}+\frac{\textstyle 7}{\textstyle
2}\beta_{1}\beta_{2}d_{2}. \notag
\end{split}
\end{equation}
Here $\beta_{j}=(4\pi)^{j+1}b_{j},~j=0-4$, $b_{j}$ are
coefficients of $\beta$-functions in $j$-loop approach [2] with
that $\beta_{4}=0$ because of $b_{4}$ is unknown [87]. Scheme
described above for DM is used in this case too for choice of
$N_{f}$ depending on $\sqrt{s_{\mbox{\footnotesize {min}}}}$
value. Detail study shown, that at the fixed values of $d_{i}$
from [83] SDGM allows to obtain the description of experimental
data at a qualitative level for both considered samples only at $
\sqrt{s_{\mbox{\footnotesize{min}}}}=10$ GeV (Table
\ref{tab:5-SDGM}) with that presence of non-perturbative term
weakly influences at fit quality. Expansion of fitted range down
to the lower energies leads to significant discrepancy of fit
curve with experimental data.

\begin{figure*}
\begin{center}
\includegraphics[width=17.5cm]{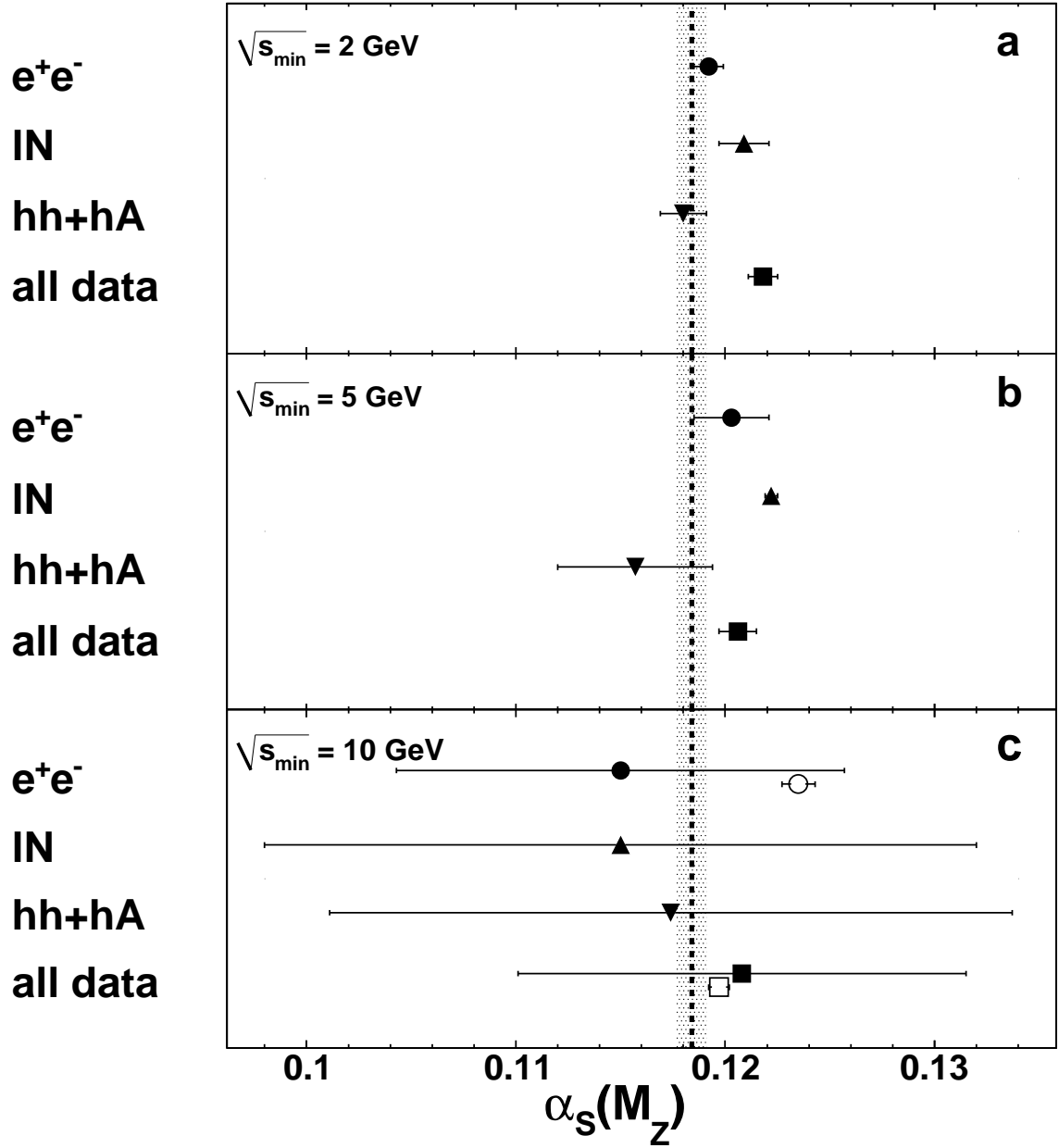}
\end{center}
\caption{Values of $\alpha_{S}(M_{Z})$ obtained by fitting of
various samples of experimental data at
$\sqrt{s_{\mbox{\footnotesize{min}}}}=2$ (a), 5 (b) and 10 GeV
(c). Values extracted in the framework of DM are shown by closed
symbols, for SDGM -- by opened ones. Dashed line and shared band
are shown the world average value with errors
$\alpha_{S}^{\mbox{\scriptsize{wa}}}(M_{Z})=0.1184 \pm 0.0007$ [2,
88].}\label{fig:1Dop}
\end{figure*}

Fig.\ref{fig:1Dop} shows values of $\alpha_{S}(M_{Z})$ obtained by
fit of dependence $\langle T\rangle (s)$ in the framework of DM
and SDGM at various $\sqrt{s_{\mbox{\footnotesize {min}}}}$.
Experimental values of $\alpha_{S}(M_{Z})$ derived at
$\sqrt{s_{\mbox{\footnotesize{min}}}}=2$ and 5 GeV for some
interactions as well as for joint data sample agree reasonably
with world average [2, 88]. Precision for estimations of strong
coupling constant considerably deteriorates at
$\sqrt{s_{\mbox{\footnotesize{min}}}}=10$ GeV in comparison with
smaller values of the low boundary of fitted energy range in the
framework of DM (Fig.\ref{fig:1Dop}c). At this value of
$\sqrt{s_{\mbox{\footnotesize{min}}}}$ SDGM allows to improve
significantly the precision of $\alpha_{S}(M_{Z})$ estimation. The
value of $\alpha_{S}(M_{Z})$ extracted in the framework of SDGM
for the joint sample is in good agreement with world average value
$\alpha_{S}^{\mbox{\scriptsize{wa}}}(M_{Z})$. On the other hand
there is some excess of experimental value of strong coupling
constant for $e^{+}e^{-}$ sample with respect to the
$\alpha_{S}^{\mbox{\scriptsize{wa}}}(M_{Z})$. However
$\alpha_{S}(M _{Z})$ experimental value shown in Table
\ref{tab:5-SDGM} for $e^{+}e^{-}$ coincides within errors with the
results obtained, for example, at investigation of radiative
decays of heavy quarkonium ($\Upsilon$) and shape of final
hadronic state in ALEPH experiment [2]. Thus, investigation of
energy dependence of average thrust values for various
interactions allows to extract estimations of $\alpha_{S}(M_{Z})$
which agree reasonably both with world average and with results of
other methods.

\begin{figure*}
\begin{center}
\includegraphics[width=17.5cm]{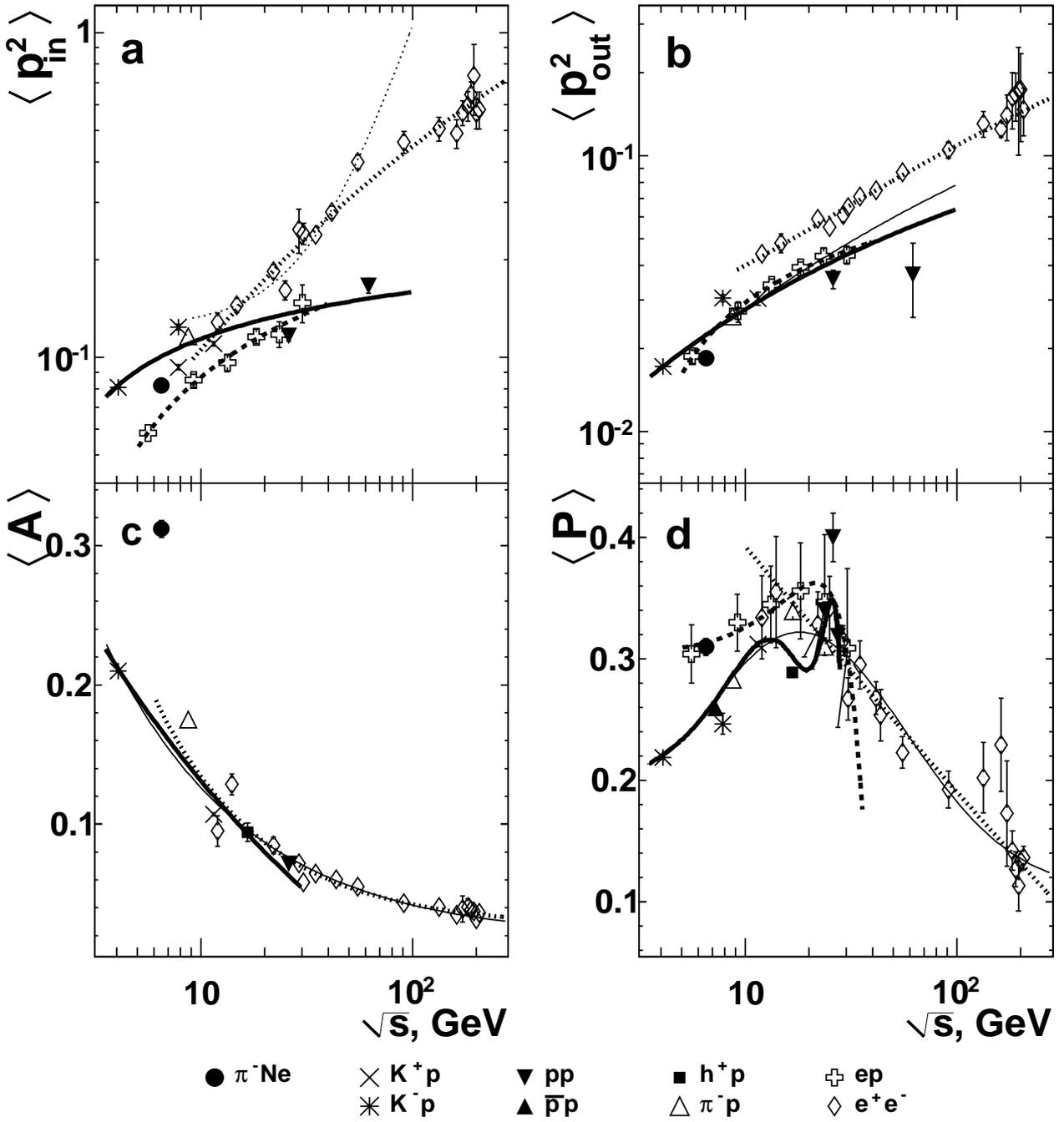}
\end{center}
\caption{Average values of the parameters describing the event
shape in the "principal axes" coordinate system depend on initial
energy: a -- $\langle
p_{\mbox{\footnotesize{\,in}}}^{\,2}\rangle$, b -- $\langle
p_{\mbox{\footnotesize{\,out}}}^{\,2}\rangle$, c -- $\langle
\mbox{A}\rangle $, d -- $\langle \mbox{P}\rangle$. Values of
$\langle p_{\mbox{\footnotesize{\,in}}}^{\,2}\rangle$ and $\langle
p_{\mbox{\footnotesize{\,out}}}^{\,2}\rangle$ are shown in
(GeV/$c$)$^{2}$. Experimental results and estimations are from
database presented in Table \ref{tab:2}. Total errors are shown.
See text for explanation of the curves.}\label{fig:2}
\end{figure*}

Figure \ref{fig:2} shows energy dependence for $\langle
p_{\mbox{\footnotesize{\,in}}}^{\,2}\rangle$ (a), $\langle
p_{\mbox{\footnotesize{\,out}}}^{\,2}\rangle$ (b), $\langle
\mbox{A}\rangle$ (c) and $\langle \mbox{P}\rangle$ (d).
Experimental results and the estimations for collective
observables are taken from a database presented in Table
\ref{tab:2}. Dependencies for all collective parameters under
study show smooth behavior, without qualitative changes at
transition from intermediate energy domain to high energies
(Fig.\ref{fig:2}). Unlike the situation for one of the main
collective parameters, $T$, the amount of phenomenological models
describing behavior of observables depending on initial energy
considered on Fig.\ref{fig:2}, is essential less or they are
absent at all. Therefore in the framework of present paper the
wide set of functions has been studied for empirical
approximations of experimental results for total available energy
range.

Experimental results for hadron-nuclear $\pi^{-}\mbox{Ne}$
interactions agree both with results for hadron-hadron reactions
at similar energies and with general trends reasonably in case of
a component of a transverse momentum in "principal axes" frame
(Fig.\ref{fig:2}a and Fig.\ref{fig:2}b). On the other hand the
multinucleon $\pi^{-}\mbox{Ne}$ collisions show essentially larger
values of parameters $\langle \mbox{A}\rangle $ (Fig.\ref{fig:2}c)
and $\langle \mbox{P}\rangle$ (Fig.\ref{fig:2}d), i.e.
hadron-nuclear collisions at $\sqrt{s} \simeq 6$ GeV are
characterized by weaker alignment of hadrons with respect to the
jet axis and smaller planarity of events in comparison with
hadronic interactions, that can be explained, possibly, by
rescattering of hadronic states in nuclear matter discussed above.
Therefore the experimental results for $\pi^{-}\mbox{Ne}$ are
excluded from the fit samples in the case of $\langle
\mbox{A}\rangle $ (Fig.\ref{fig:2}c) and $\langle \mbox{P}\rangle$
(Fig.\ref{fig:2}d) variables below.
\begin{table*}
\caption{Fit results for energy dependence of collective
observables under considered} \label{tab:4}
\begin{center}
\begin{tabular}{cccccc} \hline
\multicolumn{1}{c}{Data} & \multicolumn{1}{c}{Function} &
\multicolumn{4}{c}{Fit parameter}\\ \cline{3-6}
sample & & $a_{1}$ & $a_{2}$ & $a_{3}$ &$\chi^{2}/\mbox{ndf}$ \rule{0pt}{14pt}\\
\hline
\multicolumn{6}{c}{$\langle p_{\mbox{\footnotesize{\,in}}}^{\,2}\rangle$} \rule{0pt}{12pt} \\
\hline
$\mbox{hh}$ and $\mbox{hA}$ & (\ref{eq:Fit-vs-s-1})&$-36 \pm 28$ & $36 \pm 28$ & $(1.8 \pm 1.4) \times 10^{-3}$ & 51 \rule{0pt}{12pt}\\
$e\mbox{p}$               & --//--&$-0.08 \pm 0.03$     & $0.06 \pm 0.02$     & $0.7 \pm 0.3$     & 0.53\\
$e^{+}e^{-}$              & --//--&$0.033 \pm 0.025$ & $(1.6
\pm 0.8) \times 10^{-3}$ & $2.5 \pm 0.3$ & 2.24 \\
\hline
\multicolumn{6}{c}{$\langle p_{\mbox{\footnotesize{\,out}}}^{\,2}\rangle$} \rule{0pt}{12pt}\\
\hline
$\mbox{hh}$ and $\mbox{hA}$ & (\ref{eq:Fit-vs-s-1}) &$0.007 \pm 0.003$ & $(2.1 \pm 0.8) \times 10^{-3}$ & $1.5 \pm 0.3$ & 31 \rule{0pt}{12pt}\\
$e\mbox{p}$               & --//--&$-0.119 \pm 0.009$                 & $0.099 \pm 0.007$        & $0.26 \pm 0.02$ & 0.19\\
$(\mbox{hh,~hA})$+$e\mbox{p}$ & --//--&$(1.08 \pm
0.02) \times 10^{-2}$ & $(1.6 \pm 0.04)\times10^{-3}$ & $1.0$ (fixed) & 12 \\
$e^{+}e^{-}$              & --//--&$0.024 \pm 0.007$ & $(4 \pm 2) \times 10^{-4}$ & $2.4 \pm 0.3$ & 0.79 \\
\hline
\multicolumn{6}{c}{$\langle \mbox{A}\rangle$} \rule{0pt}{12pt}\\
\hline
$\mbox{hh}$  & (\ref{eq:Fit-vs-s-2})&$-0.19 \pm 0.07$  & $0.57 \pm 0.06$   & $-0.12 \pm 0.03$   & 195 \rule{0pt}{14pt}\\
$e^{+}e^{-}$ & --//--&$0.027 \pm 0.004$ & $0.7 \pm 0.2$     & $-0.42 \pm 0.06$   & 2.29 \\
all data   & --//--&$0.017 \pm 0.002$ & $0.483 \pm 0.008$ & $-0.323 \pm 0.008$ & 25.3 \\
\hline
\multicolumn{6}{c}{$\langle \mbox{P}\rangle$} \rule{0pt}{12pt}\\
\hline
$e\mbox{p}$  & (\ref{eq:Fit-vs-s-3})&$0.30 \pm 0.02$  & $(2.6 \pm 1.4) \times 10^{-4}$ & $(-2.8\pm1.7)\times10^{-7}$   & 0.16 \rule{0pt}{14pt}\\
$e^{+}e^{-}$ & (\ref{eq:Fit-vs-s-2})&$-3.24 \pm 0.07$ &  $3.85 \pm 0.07$ & $-0.012 \pm 0.001$   & 0.88 \\
\hline
\end{tabular}
\end{center}
\end{table*}

Comparison of Fig.\ref{fig:2}a with Fig.\ref{fig:2}b demonstrates
that value of "in"-component of transverse momentum is larger than
corresponding "out"-component value at all considered initial
energy values $\sqrt{s} \geq 4$ GeV and for all interaction types.
In the energy domain accessible for comparison, values of $\langle
p_{\mbox{\footnotesize{\,in}}}^{\,2}\rangle$ (Fig.\ref{fig:2}a)
and $\langle p_{\mbox{\footnotesize{\,out}}}^{\,2}\rangle$
(Fig.\ref{fig:2}b) show faster growth at increasing of $\sqrt{s}$
in case of $e^{+}e^{-}$ annihilation in comparison with
interactions of other types. It is agreed with results of the
comparative analysis executed in [60] for mean square of
transverse momentum, $\langle
p_{\footnotesize{\,\perp}}^{\,2}\rangle$, at $10 < \sqrt {s} < 35$
GeV. In the framework of present paper energy dependence for first
moments of distributions of "in"- and "out"-component of
transverse momentum is fitted by function
\begin{equation}
\langle p_{\footnotesize{\,j}}^{\,2}\rangle (s)=a_{1}+a_{2}[\ln
s]^{a_{3}},~~~j \equiv \mbox{in},~\mbox{out}.\label{eq:Fit-vs-s-1}
\end{equation}
Here $s$ is in GeV$^{2}$. Results of fit of components of
transverse momentum by function (\ref{eq:Fit-vs-s-1}) are
presented in Table \ref{tab:4} for various experimental data
samples and shown on Fig.\ref{fig:2}a,b for $\mbox{hh}$ and
$\mbox{hA}$ collisions by a solid line, for $e\mbox{p}$ -- by
dashed line and for $e^{+}e^{-}$ annihilation -- by dotted one.
The behavior of $\langle
p_{\mbox{\footnotesize{\,in}}}^{\,2}\rangle (s)$ and $\langle
p_{\mbox{\footnotesize {\,out}}}^{\,2}\rangle (s)$ depends on type
of interaction, especially visible for the first observable.
Disorder of experimental points and small errors do not allow to
get reasonable fit quality for hadronic interactions, however
function (\ref{eq:Fit-vs-s-1}) agrees with the general tendency of
experimental results at a qualitative level. Usually experimental
data and estimations for $e^{+}e^{-}$ and $e\mbox{p}$ are
described by the offered function (\ref{eq:Fit-vs-s-1}) with
statistically acceptable quality both for $\langle
p_{\mbox{\footnotesize{\,in}}}^{\,2}\rangle$ (Fig.\ref{fig:2}a),
and for $\langle p_{\mbox{\footnotesize{\,out}}}^{\,2}\rangle$
(Fig.\ref{fig:2}b). As it has been stressed above, the estimations
derived on the basis of published $e^{+}e^{-}$ experimental data
are used for, in particular, $\langle
p_{\mbox{\footnotesize{\,in}}}^{\,2}\rangle$ and $\langle
p_{\mbox{\footnotesize{\,out}}}^{\,2}\rangle$ at $\sqrt{s} >
M_{Z}$. Direct experimental measurements are available for these
observables up to $\sqrt{s}=55.2$ GeV in $e^{+}e^{-}$ annihilation
[21]. Therefore results for $e^{+}e^{-}$ have been studied in
energy range $\sqrt{s} < 60$ GeV separately, in which $\langle
p_{\mbox{\footnotesize{\,in}}}^{\,2}\rangle$ and $\langle
p_{\mbox{\footnotesize{\,out}}}^{\,2}\rangle$ obtained from
experiment data directly. For parameter $\langle
p_{\mbox{\footnotesize{\,in}}}^{\,2}\rangle$ better fit quality
turns out for power function (\ref{eq:Fit-vs-s-2}) at $Y \equiv
p_{\mbox{\footnotesize{\,in}}}^{\,2}$ in case under considered.
Fit function for energy range indicated above is shown on
Fig.\ref{fig:2}a by thin dotted line, values of fit parameters are
following: $a_{1}=0.124 \pm 0.005,~a_{2} =(9.3 \pm 0.6) \times
10^{-5},~a_{2} =1.0~(\mbox{fixed}),~\chi^{2}/\mbox{ndf}= 1.62$.
Thus, measurements only at $\sqrt{s} < 60$ GeV assume
significantly faster growth of $\langle
p_{\mbox{\footnotesize{\,in}}}^{\,2}\rangle$, than that observed
at the account of estimations at LEP energies. For $\langle
p_{\mbox{\footnotesize {\,out}}}^{\,2}\rangle$ the published
results and estimations at collision energies larger than $M_{Z}$
show very smooth behavior without any features. Values of fit
parameter for (\ref{eq:Fit-vs-s-1}) are coincide within errors for
fitting of all energy range and for $\sqrt{s} < 60$ GeV domain
only. Fitting by function (\ref{eq:Fit-vs-s-1}) has been made for
joint sample of results for hadronic and $e\mbox{p}$ interactions
because of close values of $\langle p_{\mbox{\footnotesize
{\,out}}}^{\,2}\rangle$ for these collisions. The values of fit
parameters presented in Table \ref{tab:4}, close within errors
with results for $\mbox {hh}+\mbox {hA}$ data sample.

Alignment of secondary particles with respect to the jet axis
amplifies with increasing of collision energy both in hadronic and
in $e^{+}e^{-}$ interactions (Fig.\ref{fig:2}c). In the framework
of present paper dependence $\langle \mbox{A}\rangle (s)$ is
fitted by function (\ref{eq:Fit-vs-s-2}), where $Y \equiv
\mbox{A}$ in case under study. Fit results for various samples of
experimental data are presented Table \ref{tab:4} and shown on
Fig.\ref{fig:2}c by solid line for $\mbox{hh}$ interactions, for
$e^{+}e^{-}$ annihilation -- by dotted one and for sample of all
available data -- by thin solid curve. As seen from
Fig.\ref{fig:2}c the behavior of collective observable $\langle
\mbox{A}\rangle$ depending on initial energy is described by
function (\ref{eq:Fit-vs-s-2}) for all sample under study at a
qualitative level.

Absence of experimental data in the range of high / low energies
for $(\mbox{hh},~e\mbox{p})$ / $e^{+}e^{-}$ interactions
respectively allows to propose only qualitative hypothesis
concerning of behavior of planarity parameter depending on initial
energy. Observable growth of $\langle \mbox{P}\rangle$ up to
$\sqrt{s} \sim 30$ GeV evidences on broadening jets in a
transverse plane, that, probably, is dominated by the contribution
of soft particles and more isotropic distributions of particle
momenta in a transverse plane. At increasing of initial energy
events become more and more planar (Fig.\ref{fig:2}d). In the
present work dependence of planarity parameter on collision energy
for $e\mbox {p}$ interactions is approximated by a polynomial
function
\begin{equation}
\langle
\mbox{P}\rangle=\sum\limits^{N}_{k=0}a_{k}s^{k}\label{eq:Fit-vs-s-3}
\end{equation}
of 2-d order and by function (\ref{eq:Fit-vs-s-2}) -- for
$e^{+}e^{-}$ annihilation. Results for fitting of these
experimental data samples are presented in Table \ref{tab:4} and
shown on Fig.\ref{fig:2}d by dashed curve for $e\mbox{p}$
interactions and by dotted line for $e^{+}e^{-}$ annihilation. As
fitting function of sample of experimental data for $\mbox{hh}$
interactions had been considered power function
(\ref{eq:Fit-vs-s-2}), predicting smooth growth of planarity in
collisions of the specified type at high energies, and polynomial
function (\ref{eq:Fit-vs-s-3}) of various orders (in the range of
$N=2-8$). Better quality of approximation among the studied
functions is observed for function (\ref{eq:Fit-vs-s-3}) at $N=5$
($\chi^{2}/\mbox{ndf}=11.9$) for which following values of
parameters have been obtained: $a_{1}=0.192 \pm 0.002,~a_{2}=(1.84
\pm 0.07) \times 10^{-3},~a_{3}=(-9.1 \pm 0.5) \times
10^{-6},~a_{4}=(1.71 \pm 0.11) \times 10^{-8},~a_{5}=(-1.05 \pm
0.08) \times 10^{-11}$. This fit by function (\ref{eq:Fit-vs-s-3})
for $\mbox{hh}$ interaction sample shown on Fig.\ref{fig:2}d by
solid line, demonstrate fast decreasing of $\langle
\mbox{P}\rangle$ at $\sqrt{s} > 30$ GeV. The reason of such
behavior, as well as in case of $e\mbox{p}$ collisions, possibly,
is absence of experimental data at $\sqrt{s} > 30$ GeV. Thus,
additional experimental data are necessary for the quantitative
analysis and more unambiguous conclusion concerning the behavior
of $\langle \mbox{P}\rangle (s)$ in hadronic interactions at high
energies. During investigation of $\langle \mbox{P}\rangle$ for
all domain $\sqrt{s} > 2$ GeV different versions of additional
term in (\ref{eq:Fit-vs-s-2}) have been examined, in particular,
Landau, normal and log normal functions. These researches have
shown, that joint sample of experimental results for
$\mbox{hh},~e\mbox{p}$ and $e^{+}e^{-}$ collisions supposes fit by
the function
\begin{equation}
\langle \mbox{P}\rangle=a_{1}+a_{2}s^{a_{3}}+\frac{\textstyle
a_{4}}{\textstyle s}\exp\left[-\frac{\textstyle 1}{\textstyle
2}\left(\frac{\textstyle \ln s-a_{5}}{\textstyle
a_{6}}\right)^{2}\right] \label{eq:Fit-vs-s-4}
\end{equation}
with the best $\chi^{2}/\mbox{ndf}=5.36$ among studied functions
and with the following values of fit parameters: $a_{1}=0.11 \pm
0.02,~a_{2} =2.3 \pm 1.3,~a_{3}=-1.8 \pm 0.9,~a_{4}=-1043 \pm
587,~a_{5}=11.2 \pm 0.9,~a_{6}=2.3 \pm 0.2$. As seen from
Fig.\ref{fig:2}d the suggested approximation function shown by
thin solid line agrees qualitatively with experimental results at
intermediate energies and corresponds to expected amplification of
event planarity in high energy domain.
\begin{table*}
\caption{Estimations for event shape observables$^{*}$}
\label{tab:5}
\begin{center}
\begin{tabular}{ccccccc} \hline
\multicolumn{1}{c}{$\sqrt{s},$} & \multicolumn{6}{c}{Observable}\\
\cline{2-7}
 TeV & $\langle S\rangle$ & $\langle T\rangle$ & $\langle
 p_{\mbox{\footnotesize{\,in}}}^{\,2}\rangle$, (GeV/$c$)$^{2}$
 & $\langle p_{\mbox{\footnotesize{\,out}}}^{\,2}\rangle$, (GeV/$c$)$^{2}$& $\langle \mbox{A}\rangle$ &
 $\langle \mbox{P}\rangle$ \rule{0pt}{14pt} \\
\hline
\multicolumn{7}{c}{$e^{+}e^{-}$ annihilation} \rule{0pt}{14pt}\\
\hline
0.5 & $0.050 \pm 0.002$ & $0.955 \pm 0.005$ & $0.9 \pm 0.8$ & $0.20 \pm 0.19$ & $0.026 \pm 0.002$ & $0.11 \pm 0.02$ \\
0.8 & $0.044 \pm 0.002$ & $0.959 \pm 0.005$ & $1.1 \pm 0.9$ & $0.23 \pm 0.23$ & $0.024 \pm 0.002$ & --//--\\
1.0 & $0.042 \pm 0.002$ & $0.960 \pm 0.005$ & $1.2 \pm 1.0$ & $0.25 \pm 0.25$ & $0.023 \pm 0.002$ & --//-- \\
3.0 & $0.033 \pm 0.002$ & $0.966 \pm 0.004$ & $1.7 \pm 1.5$ & $0.35 \pm 0.37$ & $0.020 \pm 0.002$ & --//-- \\
5.0 & $0.031 \pm 0.002$ & $0.968 \pm 0.004$ & $2.0 \pm 1.8$ & $0.40 \pm 0.44$ & $0.019 \pm 0.002$ & --//-- \\
\hline
\multicolumn{7}{c}{$e\mbox{p}$ interactions} \rule{0pt}{14pt}\\
\hline
0.8 & --                & --                & $0.25 \pm 0.25$ & $0.08 \pm 0.02$ & --                & -- \\
1.6 & $0.038 \pm 0.002$ & $0.963 \pm 0.005$ & $0.28 \pm 0.27$ & $0.08 \pm 0.02$ & $0.021 \pm 0.002$ & $0.11 \pm 0.02$ \\
\hline
\multicolumn{7}{c}{$\mbox{pp}$ collisions} \rule{0pt}{14pt}\\
\hline
0.2 & $0.067 \pm 0.003$ & $0.946 \pm 0.007$ & $0.17 \pm 0.16$ & $0.08 \pm 0.05$ & $0.033 \pm 0.002$ & $0.14 \pm 0.03$ \\
0.5 & --                & --                & $0.18 \pm 0.18$ & $0.10 \pm 0.07$ & --                & -- \\
7.0 & $0.029 \pm 0.002$ & $0.969 \pm 0.004$ & $0.21 \pm 0.20$ & $0.16 \pm 0.13$ & $0.019 \pm 0.002$ & $0.11 \pm 0.02$ \\
14  & $0.027 \pm 0.002$ & $0.971 \pm 0.004$ & $0.21 \pm 0.21$ & $0.17 \pm 0.15$ & $0.018 \pm 0.002$ & --//-- \\
28  & $0.025 \pm 0.002$ & $0.973 \pm 0.003$ & $0.22 \pm 0.21$ & $0.19 \pm 0.17$ & --//-- & --//-- \\
42  & $0.024 \pm 0.002$ & $0.974 \pm 0.003$ & --//--          & $0.20 \pm 0.18$ & --//-- & --//-- \\
100 & $0.023 \pm 0.002$ & $0.976 \pm 0.003$ & $0.22 \pm 0.22$ & $0.23 \pm 0.21$ & --//-- & --//-- \\
200 & --//--            & $0.977 \pm 0.003$ & $0.23 \pm 0.22$ & $0.25 \pm 0.23$ & --//-- & --//-- \\
\hline \multicolumn{7}{l}{$^*$\rule{0pt}{11pt}\footnotesize Fit
results for joint data samples are used to estimate the $\langle
S\rangle,~\langle T\rangle,~\langle \mbox{A}\rangle$ and $\langle
\mbox{P}\rangle$ for a certain type
of interaction,} \\
\multicolumn{7}{l}{\footnotesize thus the some estimation is
indicated once if various interactions are studied at the same
energy.}
\end{tabular}
\end{center}
\end{table*}

On the basis of the above experimental results, estimations and
the suggested approximations it is possible to draw an
intermediate conclusion, that all traditional collective variables
show smooth dependence on energy without any features at $\sqrt{s}
\sim 3-5$ GeV. In Table \ref{tab:5} estimations are shown for
parameters of event geometry in "principal axes" coordinate system
in various interactions for the present and possible future
accelerating facilities, calculated in the framework of the paper
on the basis of analytical functions (\ref{eq:Fit-vs-s-2}),
(\ref{eq:4.4.Fit-vs-s-Tgeneral}) -- (\ref{eq:4.4.Fit-vs-s-TSDGM}),
(\ref{eq:Fit-vs-s-1}) -- (\ref{eq:Fit-vs-s-4}) and Table
\ref{tab:4-sf} -- \ref{tab:4}. Additional experimental results are
essential for improving of precision of estimations for $\langle
p_{\mbox{\footnotesize{\,in}}}^{\,2}\rangle$ and $\langle
p_{\mbox{\footnotesize{\,out}}}^{\,2}\rangle$. But it would be
emphasize, that the estimation of absolute value of transverse
particle momentum with respect to the jet axis $\sqrt{\langle
p_{\footnotesize{\,\perp}}^{\,2}\rangle}=0.50 \pm 0.17$ GeV/$c$
derived in the framework of present paper for $\mbox{pp}$
collisions at $\sqrt{s}=200$ GeV coincides within errors with
experimental result at the same energy [89]. Study the high and
ultra high energy domain show that the asymptotic values within
errors are predicted for most collective parameters under study at
$\sqrt{s} \sim 5-7$ TeV already.

\section{Multiplicity dependence}\label{sec:5}
Unlike energy dependence, the investigations of geometry of a
hadronic final state at the fixed secondary particle
multiplicities are less extensive and have been made for energy
range $ \sqrt {s} <60 $ GeV. Hadron-hadron interactions were
studied in the exclusive approach in the some cases for
intermediate energies. Results for collective observables were
obtained only for jets of all identified secondary particles in
some experiments (see, for example, [22]). Moreover using of total
multiplicity ($N _ {\mbox {\footnotesize {tot}}} $) allows to
increase amount of experimental points and hence to investigate
behavior of dependence of jet characteristics on multiplicity in
more detail. Therefore the event shape of a hadronic final state
is studied depending on multiplicity of identified particles
($N$), which is equal $N=N_{\mbox {\footnotesize {tot}}}$ for the
exclusive approach and $N=N_{\mbox {\footnotesize {ch}}}$ in other
cases.

\begin{figure*}
\begin{center}
\includegraphics[width=17.5cm]{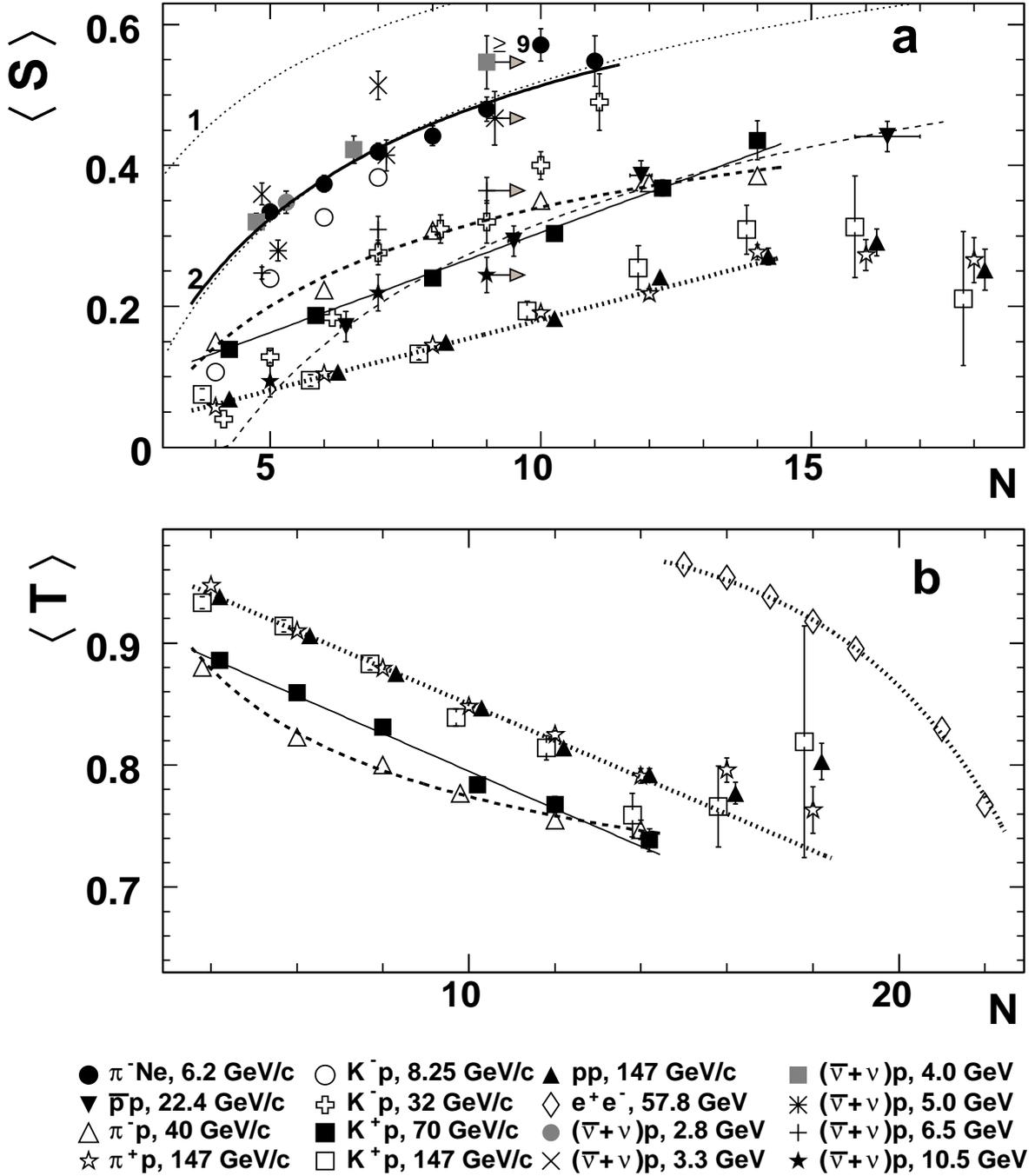}
\end{center}
\caption{Dependence of average values of sphericity (a) and thrust
(b) on multiplicity of identified particles in event. Experimental
results for $\pi^{-}\mbox{Ne}$ are from [70, 71], for
$K^{-}\mbox{p}$ -- from [22, 69], for $\pi^{-}\mbox{p}$ at 40
GeV/$c$ -- from [23], for $K^{+}\mbox {p}$ at 70 GeV/$c$ -- from
[56], for $\bar{\mbox{p}} \mbox{p}$ at 22.4 GeV/$c$ -- from [67],
for $\mbox{h}^{+}\mbox{p}$ at 147 GeV/$c$ -- from [63], for $(\bar
{\nu}+\nu)\mbox{p}$ -- from [48]. Estimations for an average
thrust values in case of $e^{+}e^{-}$ are obtained on the basis of
results from [90]. For $(\bar{\nu}+\nu)\mbox{p}$ average invariant
mass of hadronic final state, $\langle W\rangle$, is indicated
above, for $e^{+}e^{-}$ -- initial energy. Dotted thin lines
marked as "1" and "2" at (a) are phenomenological approximations
from [15]. Fit results from other publications, namely, for
sphericity (a) and thrust (b) in $\mbox{h}^{+}\mbox{p}$ hadronic
interactions from [63]; for $e^{+}e^{-}$ annihilation for thrust
(b) from [90], are shown by thick dotted lines. Statistical errors
are shown. See text for explanation of fit curves in more
detail.}\label{fig:3}
\end{figure*}

Average values of main collective variables are presented at
Fig.\ref{fig:3} depend on multiplicity of the identified secondary
particles in event. Experimental data are taken from [22, 23, 48,
56, 63, 67, 69 - 71, 90]. At $\langle W\rangle \geq 4.0$ GeV right
points correspond to integrated values $\langle S\rangle$ for a
multiplicity range with low boundary indicated for $\langle
W\rangle=4.0$ GeV namely (\ref {fig:3}a). We used the results for
modified sphericity tensor with reduced momenta of final state
particles [16] and sphericity definition from original papers for
$K^{-}\mbox{p}$ [22, 69]. Fit results of present work and
corresponding curves at Fig.\ref{fig:3} are described below. The
intensification of spherical event shape is observed with final
state multiplicity increasing for all types of interactions under
study at intermediate energy domain $\sqrt{s} < 60$ GeV
(Fig.\ref{fig:3}). Rescattering of hadronic states in nuclear
matter results in to the some increasing of $\langle S\rangle$ for
$\pi^{-}\mbox{Ne}$ in comparison with elementary reactions.
Therefore one can see the good agreement of results obtained for
$\pi^{-}\mbox{Ne}$ and for $(\bar{\nu}+\nu)\mbox{p}$ in despite of
some difference in the initial energies. Similar accordance with
general tendency is observed for $\pi^{-}\mbox{p}$,
$K^{+}\mbox{p}$, $\bar{\mbox{p}}\mbox{p}$ and
$(\bar{\nu}+\nu)\mbox {p}$ interactions at larger initial energies
too. In [15] analytical dependence $\langle S\rangle(N) \propto
N^{-0.5}$ was derived for isotropic distribution of secondary
particles in phase space for two ultimate cases:
non-re\-la\-ti\-vis\-tic and extremely relativistic ones which are
shown at Fig.\ref{fig:3}a by dotted thin curves "1" and "2"
respectively. Results for $\pi^{-}\mbox{Ne}$ and for
$(\bar{\nu}+\nu)\mbox{p}$ agree well with phenomenological
dependence for ultra-relativistic case, that confirms the
difficulty of separation of jet events at $\sqrt{s} \sim 3-5$ GeV
on the basis of one event shape observable (\ref{eq:2.Spherisity})
for interactions of any types. Thus, one can suggest, that the
universal estimation of the low energy boundary for experimental
appearance of event jet structure in multiparticle production
processes is $\sqrt{s_{c}} \sim 3$ GeV. At increasing of initial
energy one can see that estimations extracted for isotropic
one-particle distribution are essentially larger than experimental
values $\langle S \rangle$, that corresponds to an amplifying for
appearance of event jet structure. Functional behavior of
dependence $\langle T\rangle(N)$ in case of $e^{+}e^{-}$
annihilation at significantly larger initial energy qualitatively
distinguishes from behavior of $\langle T\rangle(N)$ for
hadron-hadron collisions at $\sqrt {s} < 12$ GeV but $\langle
T\rangle$ decreases with multiplicity increasing for all
interaction types under study (Fig.\ref{fig:3}b). One need to
emphasize that universality of geometry of events can be
considered as evidence for similarity of dynamics of soft hadronic
jet production in various interactions in wide initial energy
range. Therefore dependencies $\langle S\rangle(N)$ and $\langle T
\rangle(N)$ indicate at possible experimental appearance of color
degrees of freedom in hadronic jet production at $\sqrt {s} \sim
3-5$ GeV.

\begin{figure*}
\begin{center}
\includegraphics[width=17.5cm]{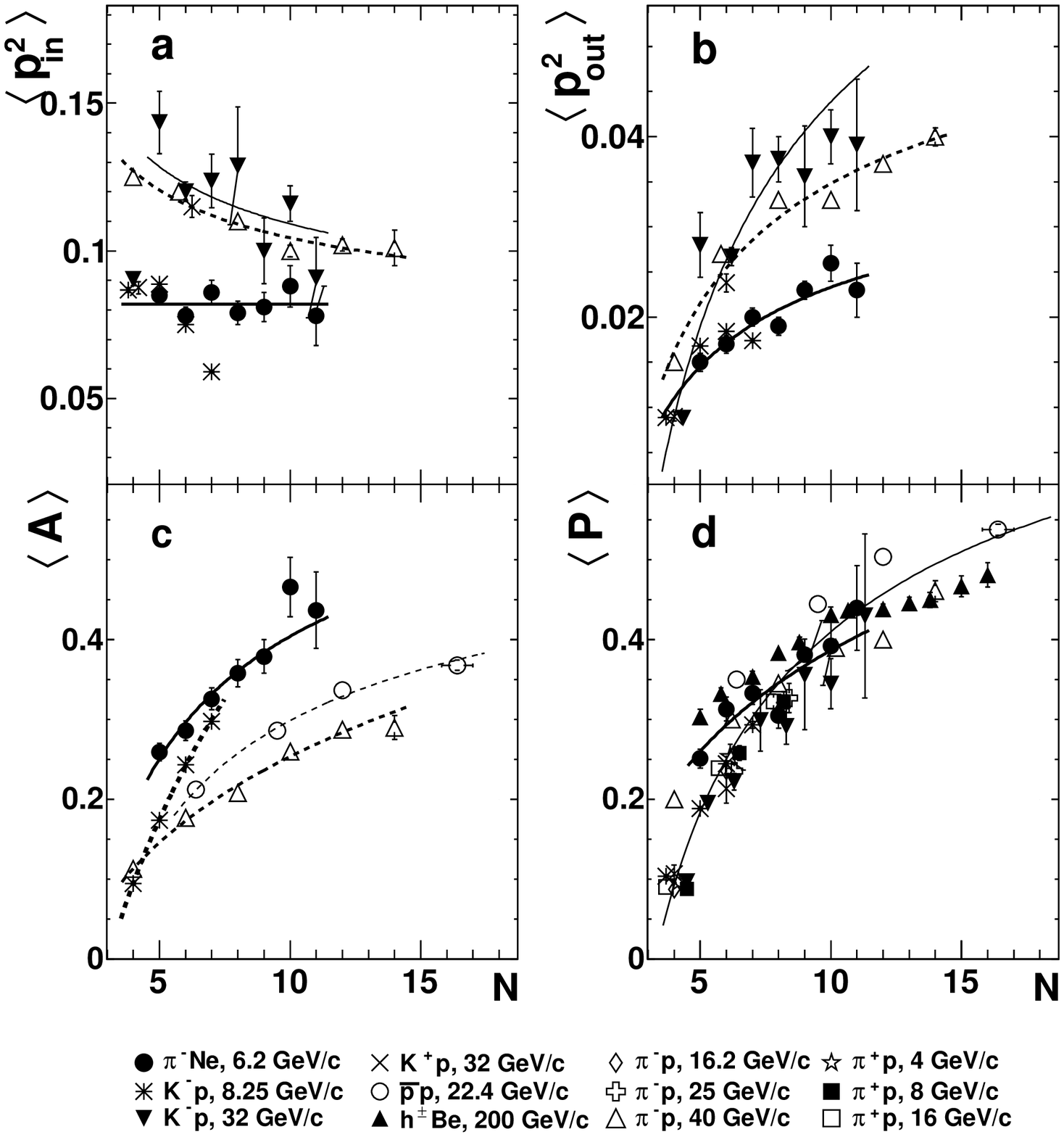}
\end{center}
\caption{Average values of collective parameters describing the
event shape in the "principal axes" coordinate system depend on
multiplicity of the secondary identified particles: a -- $\langle
p_{\mbox{\footnotesize{\,in}}}^{\,2}\rangle$, b -- $\langle
p_{\mbox{\footnotesize{\,out}}}^{\,2}\rangle$, c -- $\langle
\mbox{A}\rangle $, d -- $\langle \mbox{P}\rangle$. Values of
$\langle p_{\mbox{\footnotesize{\,in}}}^{\,2}\rangle$ and $\langle
p_{\mbox{\footnotesize{\,out}}}^{\,2}\rangle$ are shown in
(GeV/$c$)$^{2}$. Results for $\pi^{-}\mbox{Ne}$ are from [70, 71],
for $\pi^{-}\mbox{p} $ at 16, 25 GeV/$c$ and for $\pi^{+}\mbox{p}$
at 4, 8, 16 GeV/$c$ are from [54], for $K^{-}\mbox{p}$ -- from
[22, 69], for $\bar{\mbox{p}}\mbox{p}$ at 22.4 GeV/$c$ -- from
[67], for $K^{+}\mbox{p}$ at 32 GeV/$c$ -- from [68], for
$\pi^{-}\mbox{p}$ at 40 GeV/$c$ -- from [23], for
$\mbox{h}^{\pm}\mbox{Be}$ at 200 GeV/$c$ -- from [73]. Statistical
errors are shown. See text for explanation of the
curves.}\label{fig:4}
\end{figure*}

Figure \ref{fig:4} shows the multiplicity dependence for average
values of a square of in-plane transverse momentum (a), square of
transverse momentum out of the event plane (b), alignment (c), and
planarity (d). Estimation values for $\langle
p_{\mbox{\footnotesize{\,in}}}^{\,2}\rangle$ and $\langle
p_{\mbox{\footnotesize{\,out}}}^{\,2}\rangle$ in case of
$\pi^{-}\mbox {p}$ at 40 GeV/$c$ were calculated on the basis of
experimental data for eigenvalues of tensor $S^{\alpha \beta}$ and
for $N_{\mbox {\footnotesize {ch}}}$ from [23]. For
$\pi^{\pm}\mbox{p}$ interactions at initial momenta 4 -- 25
GeV/$c$ values of alignment and planarity corresponded to
definitions used in present work (Table \ref{tab:1}) have been
estimated on the basis of experimental results from [54].

One can see that behavior of $\langle
p_{\mbox{\footnotesize{\,in}}}^{\,2}\rangle(N)$ (Fig.\ref{fig:4}a)
and $\langle p_{\mbox{\footnotesize{\,out}}}^{\,2}\rangle(N)$
(Fig. \ref{fig:4}b) depends both on initial energy and from
reaction type. Available experimental results for components of
transverse momentum under study do not allow to make conclusions
concerning presence and character of the universal analytical
dependence describing experimental data for different interactions
and / or initial energies. But one can conclude that hadronic
interactions show the decreasing of $\langle
p_{\mbox{\footnotesize{\,in}}}^{\,2}\rangle(N)$ and increasing of
$\langle p_{\mbox{\footnotesize{\,out}}}^{\,2}\rangle(N)$ with
multiplicity increasing in general.
\begin{table*}
\caption{Fit results for multiplicity dependence in hadronic
interactions} \label{tab:6}
\begin{center}
\begin{tabular}{lcccc} \hline
\multicolumn{1}{c}{Sample} & \multicolumn{4}{c}{Fit parameter}
\\\cline{2-5}
 & $a_{1}$ & $a_{2}$ & $a_{3}$ (fixed) & $\chi^{2}/\mbox{ndf}$ \rule{0pt}{14pt}\\
\hline
\multicolumn{5}{c}{$\langle S\rangle$} \rule{0pt}{14pt}\\
\hline
$\pi^{-}\mbox{Ne}$       & $0.97 \pm 0.05$ & $-1.44 \pm 0.13$ & -0.5 & 1.89 \\
$\bar{\mbox{p}}\mbox{p}$ & $0.91 \pm 0.07$ & $-1.9 \pm 0.2$ & --//-- & 0.41 \\
$\pi^{-}\mbox{p}$        & $0.680 \pm 0.009$ & $-1.07 \pm 0.03$ & --//-- & 6.33 \\
$K^{+}\mbox{p}$          & $0.021 \pm 0.009$ & $0.028 \pm 0.001$ & 1.0 & 0.63 \\
\hline
\multicolumn{5}{c}{$\langle T\rangle$} \rule{0pt}{14pt} \\
\hline
$\pi^{-}\mbox{p}$ & $0.594 \pm 0.015$ & $0.57 \pm 0.04$ & -0.5 & 0.22 \\
$K^{+}\mbox{p}$   & $0.949 \pm 0.005$ & $-0.015 \pm 0.001$ & 1.0 & 1.94 \\
\hline
\end{tabular}
\end{center}
\end{table*}

Parameters characterized of alignment (Fig.\ref{fig:4}c) and
planarity (Fig.\ref{fig:4}d) of event show smooth growth at
multiplicity increasing for all types of interactions and initial
energy domain under study, i.e. the events are less planar and are
characterized smaller alignment at larger hadron multiplicity in
final state. Decreasing of values of parameter $\langle
\mbox{A}\rangle$ is observed with increasing of beam momentum,
i.e. the alignment of hadrons along the principal axis
$\vec{n}_{1}$ (jet axis) is amplified for higher initial energies.
Influence of strange particles in a final state leads to small
values of alignment in case of $K^{-}\mbox{p}$ interactions at
initial momentum 8.25 GeV/$c$ [22] at small multiplicities.
Moreover estimations of alignment parameter for $K^{\pm}\mbox{p}$
at 32 GeV/$c$ [68, 69] are smaller significantly in the case of
sphericity tensor definition (\ref{eq:4.Tensor-Sphericity}) than
that for other experiments at close energies. Comparison of
Fig.\ref{fig:4}c and Fig.\ref{fig:4}d demonstrates, that the
parameter $\langle \mbox{P}\rangle $ is less sensitive to presence
of strange particles in final state, than alignment. Detail
additional study shows that behaviors of dependencies at
Fig.\ref{fig:3} and at Fig.\ref{fig:4} are similar for total and
charged multiplicities and the general picture changes
insignificantly at substitution $N_{\mbox {\footnotesize {tot}}}$
on $N_{\mbox {\footnotesize {ch}}}$ and vase versa.

The situation with theoretical description of behavior of
multiplicity dependence for collective parameters differs from the
situation for energy dependence of some observables, at least.
Apparently, there is no theoretical or phenomenological
predictions for multiplicity dependence beyond [15] or
experimental data fits for main collective variables [63, 90].
Taking into account monotonic and smooth behavior of dependencies
under study, experimental data for average values of collective
parameters in hadronic and (anti)neutrino interactions have been
fitted by general power function (\ref{eq:Fit-vs-s-2}) for $Y
\equiv
S,~T,~p_{\mbox{\footnotesize{\,in}}}^{\,2},~p_{\mbox{\footnotesize{\,out}}}^{\,2},~\mbox{A},~\mbox{P}$
and $X \equiv N$ in the case considered in this section. Detail
study of fit results for various data ensembles for all collective
observables has shown, that better fit quality is achieved,
usually, for a special case of (\ref{eq:Fit-vs-s-2}) at the
following fixed values $a_{3}=-0.5$ or $a_{3}=1.0$. The first case
agrees with function suggested in [15] for sphericity.
Approximations are shown by thick solid line for
$\pi^{-}\mbox{Ne}$ and by thin dashed line for
$\bar{\mbox{p}}\mbox{p}$ at 22.4 GeV/$c$ on Fig.\ref{fig:3}a; by
thick dashed line for $\pi^{-}\mbox{p}$ and by thin solid line for
$K^{+}\mbox{p}$ at 70 GeV/$c$ on Fig.\ref{fig:3}. Results for fit
by function (\ref{eq:Fit-vs-s-2}) at fixed $a_{3}$ are shown in
Table \ref{tab:6} for hadronic interactions under study. As seen,
fit quality are acceptable for most data samples. One need to
emphasize that general function (\ref{eq:Fit-vs-s-2}) allows to
get the best quality for $K^{+}\mbox{p}$ collisions
($\chi^{2}/\mbox{ndf}=0.01$) with the following values of fit
parameters: $a_{1}=0.08 \pm 0.03$, $a_{2}=0.009 \pm 0.006$,
$a_{3}=1.4 \pm 0.3$. But the $a_{3}$ value is close to unit within
errors and this general approximation agrees quite reasonable with
linear one. As indicated above, modified both sphericity tensor
and definition of corresponding observable from original papers
which slightly differs from definition indicated in Table
\ref{tab:1} are used for $K^{-}\mbox{p}$ interactions. Reducing of
influence of leading particles results in some faster
intensification of spherical event shape with multiplicity
increasing in comparison with behavior of $\langle S\rangle(N)$
dependence for other hadronic interactions at close energies for
sphericity tensor definition (\ref{eq:4.Tensor-Sphericity}).
Otherwise, results for $K^{-}\mbox{p}$ interactions at 8.25
GeV/$c$ and 32 GeV/$c$ agree with other experimental points
qualitatively but for the first case corresponding fit quality is
statistical unacceptable due to, possibly, small experimental
errors. The following results have been obtained for fit of
$\langle S\rangle(N)$ for $K^{-}\mbox{p}$ collisions at 32 GeV/$c$
by (\ref{eq:Fit-vs-s-2}): $a_{1}=-0.94 \pm 0.01$, $a_{2}=0.608 \pm
0.009$, $a_{3}=0.346 \pm 0.006$ at $\chi^{2}/\mbox{ndf}=1.33$ (we
do not show the curve at Fig.\ref{fig:3}a for more clearly
picture). Multiplicity dependence of sphericity for (anti)neutrino
beams are fitted by function (\ref{eq:Fit-vs-s-2}) with fixed
$a_{3}=-0.5$ at all energies with statistically reasonable
qualities. Results of fit for $(\bar{\nu}+\nu)\mbox{p}$
interactions are presented in Table \ref{tab:7} (we do not show
corresponding curves at Fig.\ref{fig:3}a for more clearly
picture).

Experimental data for various interaction shown at Fig.\ref{fig:3}
and fit results for main collective variables (Tables \ref{tab:6}
and \ref{tab:7}) allow to suggest that intensification of
spherical event shape with multiplicity increasing seems more
powerful in energy domain $\sqrt{s} > 11$ GeV than that at lower
initial energies.
\begin{table}[!h]
\caption{Fit results for multiplicity dependence of $\langle
S\rangle$ at (anti)neutrino beams} \label{tab:7}
\begin{center}
\begin{tabular}{lccc} \hline
\multicolumn{1}{c}{$\langle W\rangle$, GeV} &
\multicolumn{3}{c}{Fit parameter}
\\\cline{2-4}
 & $a_{1}$ & $a_{2}$ & $\chi^{2}/\mbox{ndf}$ \rule{0pt}{14pt}\\
\hline
4.0  & $1.09 \pm 0.11$ & $-1.7 \pm 0.3$ & 1.34 \\
5.0  & $1.08 \pm 0.12$ & $-1.8 \pm 0.3$ & 0.42 \\
6.5  & $0.69 \pm 0.07$ & $-1.0 \pm 0.2$ & 0.19 \\
10.5 & $0.72 \pm 0.11$ & $-1.4 \pm 0.3$ & 1.22 \\
\hline
\end{tabular}
\end{center}
\end{table}
\begin{table}[!h]
\caption{Values of fit parameters for function
(\ref{eq:Fit-vs-s-2})} \label{tab:8}
\begin{center}
\begin{tabular}{lccc} \hline
\multicolumn{1}{c}{Data} & \multicolumn{3}{c}{Fit parameter}
\\\cline{2-4}
sample & $a_{1}$ & $a_{2}$ & $\chi^{2}/\mbox{ndf}$ \rule{0pt}{14pt}\\
\hline
\multicolumn{4}{c}{$\langle p_{\mbox{\footnotesize{\,in}}}^{\,2}\rangle$} \rule{0pt}{12pt} \\
\hline
$\pi^{-}\mbox{Ne}$ & $0.082 \pm 0.002$ & $0.0$ (fixed) & ~0.88 \rule{0pt}{12pt}\\
$K^{-}\mbox{p}$ & $0.06 \pm 0.02$ & $0.15 \pm 0.06$ & 1.47 \\
$\pi^{-}\mbox{p}$ & $0.065 \pm 0.004$ & $0.123 \pm 0.012$ & 2.78\\
\hline
\multicolumn{4}{c}{$\langle p_{\mbox{\footnotesize{\,out}}}^{\,2}\rangle$} \rule{0pt}{12pt} \\
\hline $\pi^{-}\mbox{Ne}$ & $0.045 \pm 0.004$ & $-0.07
\pm 0.01$ & ~1.36 \rule{0pt}{12pt}\\
$K^{-}\mbox{p}$ & $0.104 \pm 0.004$ & $-0.190 \pm 0.008$ & 1.98 \\
$\pi^{-}\mbox{p}$ & $0.067 \pm 0.002$ & $-0.102 \pm 0.005$ & 2.82\\
\hline
\multicolumn{4}{c}{$\langle \mbox{A}\rangle$} \rule{0pt}{12pt} \\
\hline $\pi^{-}\mbox{Ne}$ & $0.77 \pm 0.06$ & $-1.17
\pm 0.14$ & ~0.89 \rule{0pt}{12pt}\\
$\bar{\mbox{p}}\mbox{p}$ & $0.65 \pm 0.01$ & $-1.10 \pm 0.03$ & 3.96 \\
$\pi^{-}\mbox{p}$ & $0.50 \pm 0.01$ & $-0.79 \pm 0.03$ & 6.15\\
\hline
\multicolumn{4}{c}{$\langle \mbox{P}\rangle$} \rule{0pt}{12pt} \\
\hline $\pi^{-}\mbox{Ne}$ & $0.68 \pm 0.06$ & $-0.94
\pm 0.15$ & 2.06 \rule{0pt}{12pt}\\
$K$-beams & $0.885 \pm 0.013$ & $-1.56 \pm 0.03$ & 2.47 \\
$\pi$-beams & $0.912 \pm 0.010$ & $-1.64 \pm 0.02$ & 8.13 \\
\hline
\end{tabular}
\end{center}
\end{table}

Notations for approximation curves at Fig.\ref{fig:4} are the same
as well as on Fig.\ref{fig:3} for corresponding reactions.
Moreover the fit curve for $K^{-}\mbox{p}$ at 32 GeV/$c$ is shown
by solid thin line on Fig.\ref{fig:4}a, b; for $K^{-}\mbox{p}$ at
8.25 GeV/$c$ -- by dotted line at Fig.\ref{fig:4}c; and curve for
all available hadronic interactions is demonstrated by thin solid
line at Fig.\ref{fig:4}d. For multiplicity dependence of alignment
in $K^{-}\mbox{p}$ reaction at 8.25 GeV/$c$ the following values
of fit parameters have been obtained for (\ref{eq:Fit-vs-s-2}):
$a_{1}=-3.035 \pm 0.008,~a_{2}=2.676 \pm 0.008,~a_{3}=0.113 \pm
0.001$, and $\chi^{2}/\mbox{ndf}=1.55$. Fit result for various
experimental data samples by specific case of function
(\ref{eq:Fit-vs-s-2}) at fixed $a_{3}=-0.5$ are shown in Table
\ref{tab:8} for all interactions and event shape variables with
the exception of $\langle
p_{\mbox{\footnotesize{\,in}}}^{\,2}\rangle (N)$ for
$\pi^{-}\mbox{Ne}$ at 6.2 GeV/$c$ discussed below. As seen,
multiplicity dependence for all collective variables are fitted by
specific case of (\ref{eq:Fit-vs-s-2}) with reasonable quality in
the case of $\pi^{-}\mbox{Ne}$ hadron-nuclear collisions. For this
reaction dependence of $\langle
p_{\mbox{\footnotesize{\,in}}}^{\,2}\rangle$ on $N$
(Fig.\ref{fig:4}a) is fitted with best quality by the simplest
case of (\ref{eq:Fit-vs-s-2}) which corresponds to constant. On
the other hand fit by functional form (\ref{eq:Fit-vs-s-2}) with
fixed $a_{3}=-0.5$ shows reasonable and only slightly poorer
quality ($\chi^{2}/\mbox{ndf}$=1.02). Thus the behavior of
$\langle p_{\mbox{\footnotesize{\,in}}}^{\,2}\rangle (N)$ does not
contradict with specific case of function (\ref{eq:Fit-vs-s-2})
indicated above for hadron-nuclear reaction under study. The
reasonable quality has been obtained for fit of data sample for
$\langle p_{\mbox{\footnotesize{\,in}}}^{\,2}\rangle$ in
$K^{-}\mbox{p}$ at 32 GeV/$c$ without first point only.
Dependencies of $\langle
p_{\mbox{\footnotesize{\,in}}}^{\,2}\rangle$ and $\langle
p_{\mbox{\footnotesize{\,out}}}^{\,2}\rangle$ on $N$ shown for
$K^{-}\mbox{p}$ at 8.25 GeV/$c$~on Fig.\ref{fig:4}a and
Fig.\ref{fig:4}b, respectively, agree with approximation function
(\ref{eq:Fit-vs-s-2}) at $a_{3}=-0.5$ qualitatively but the
corresponding fit qualities are statistical unacceptable for this
reaction due to dispersion of experimental points and, possibly,
small errors. As seen from Table \ref{tab:8}, planarity results
obtained in interactions with kaon beams are fitted by
(\ref{eq:Fit-vs-s-2}) with reasonable $\chi^{2} / \mbox{ndf}$.
Experimental results for pion beams agree with approximation
function (\ref{eq:Fit-vs-s-2}) qualitatively only and fit quality
is some poorer for this incoming particle than that for kaon
beams. The approximation curves for two beam types under study are
very close to each other, moreover results for all hadronic
interactions demonstrate similar values for planarity parameter
$\langle P\rangle$ at corresponding multiplicities
(Fig.\ref{fig:4}d). Thus we have attempted to fit all available
experimental points by (\ref{eq:Fit-vs-s-2}) at fixed
$a_{3}=-0.5$. One can conclude from Fig.\ref{fig:4}d that total
sample of experimental results for hadronic interactions agrees
with specific case of function (\ref{eq:Fit-vs-s-2}) at
qualitative level but the fit quality is statistically
unacceptable ($\chi^{2} / \mbox{ndf} \simeq 20$).

Additional investigations show the following. First of all, power
function (\ref{eq:Fit-vs-s-2}) and it's special case at fixed
$a_{3}=-0.5$ describe dependencies of collective variables on
char\-ged multiplicity reasonable at qualitative level too.
Second, as expected, using of modified sphericity tensor with
reduced momenta of final state particles for $K^{-}\mbox{p}$ [22,
69] result in growth of values of $\langle \mbox{A}\rangle$
(especially) and $\langle \mbox{P}\rangle$ with the preservation
of functional form for corresponding dependencies. On the other
hand the behavior of dependence of components of particle
transverse momentum in and out of the event plane on multiplicity
changes dramatically, namely, $\langle
p_{\mbox{\footnotesize{\,in}}}^{\,2}\rangle$ and $\langle
p_{\mbox{\footnotesize{\,out}}}^{\,2}\rangle$ are shown the power
growth with $N$ increasing and significantly larger values than
that for sphericity tensor definition
(\ref{eq:4.Tensor-Sphericity}).

\section{Summary}\label{sec:6}
\hspace*{0.5cm}In the framework of present paper the database of
experimental results for traditional event shape variables was
created for wide set of interaction types. This database is used
for investigation of dependencies of collective parameters on
initial energy and multiplicity in various interactions and for
joint samples.

At intermediate energies influence of nuclear matter, in
particular, multinucleon interactions leads to significantly more
isotropic distribution of secondary particles for hadron-nuclear
reactions in comparison with hadron-hadron ones, to growth of
transverse momentum, attenuation of alignment and planarity of
events. Leading particles result in the inverse influence on event
shape in hadronic reactions in comparison with intranuclear
rescattering.

Energy dependencies of the main collective observables sphericity
and thrust show universal behavior for various interactions.
Phenomenological models based on the QCD, in particular,
dispersive model allow to describe $\langle T\rangle (s)$ in wide
range of initial energies down to strongly non-perturbative domain
$\sqrt{s} \sim 2-3$ GeV at qualitative level at least.
Approximation of energy dependence of average thrust values for
various interactions allows to extract $\alpha_{S}(M_{Z})$
estimations which are in agreement both with world average value
and with results obtained by other methods. Values of transverse
momentum components which are in / out of event plane for
$e^{+}e^{-}$ interactions increase faster with growth of
$\sqrt{s}$, than that for $e\mbox{p}$ and hadronic interactions.
Energy dependencies of alignment and planarity suppose the
descriptions by universal empirical functions for different
interaction types. Using suggested analytic approximation
functions estimations of values of collective parameters under
study have been obtained for present and future collider-research
facilities. At $\sqrt{s} \sim 5-7$ TeV average values of
collective variables studied in the present paper do not depend on
initial energy within errors. Thus, these values of $\sqrt{s}$ can
be considered as an estimation of the low bound of asymptotic
region for traditional collective parameters.

Spherical event shape is amplified with multiplicity increasing
and one can suggest, that the universal estimation of the low
energy boundary for experimental appearance of event jet structure
in multiparticle production processes is $\sqrt{s_{c}} \sim 3$
GeV. Usually, multiplicity dependence of collective variables
under study agree with power function in energy domain $\sqrt{s} <
12$ GeV at qualitative level at least.

\end{document}